\documentclass[fleqn,usenatbib,useAMS]{mnras}

\usepackage{graphicx}	
\usepackage{amsmath}	
\usepackage{multicol}        
\usepackage{bm}		
\usepackage{pdflscape}	
\usepackage{comment}

\usepackage[T1]{fontenc}
\usepackage{ae,aecompl}
\usepackage{xcolor}

\usepackage{newtxtext,newtxmath}

\title[Signatures of Accreting Black Holes in Line Intensity Mapping]{Signatures of Black Holes in Line Intensity Mapping}

\author[E. Ogata et al.]{Erika Ogata$^{1}$\thanks{Contact e-mail: erika.ogata@phys.s.u-tokyo.ac.jp}, 
Kana Moriwaki$^{1,2}$, 
Adam Lidz$^{3}$,
Rui Lan Jun$^{2}$,
Naoki Yoshida$^{2,4,5}$%
\\
$^{1}$Research Center for the Early Universe, Graduate School of Science, The University of Tokyo, 7-3-1 Hongo, Bunkyo, Tokyo 113-0033, Japan\\
$^{2}$Department of Physics, Graduate School of Science, The University of Tokyo, 7-3-1 Hongo, Bunkyo, Tokyo 133-0033, Japan\\
$^{3}$Department of Physics and Astronomy, University of Pennsylvania 209 South 33rd Street,Philadelphia, PA 19104, USA\\
$^{4}$Kavli Institute for the Physics and Mathematics of the Universe (WPI), UT Institute for Advanced Study, The University of Tokyo, Kashiwa, Chiba 277-8583, Japan\\
$^{5}$RIKEN Center for Advanced Intelligence Project, 1-4-1 Nihonbashi, Chuo, Tokyo 103-0027, Japan\\
}

\date{Last updated 2024 April 23; in original form 2013 September 5}

\pubyear{{\the\year{}}}

\begin{document}
\label{firstpage}
\pagerange{\pageref{firstpage}--\pageref{lastpage}}
\maketitle

\begin{abstract}
Line-intensity mapping (LIM) 
has attracted growing attention as a powerful technique for probing the large-scale distribution of galaxies and the cosmic history of star formation through unresolved line emission. 
Existing LIM models for galaxy-associated lines, such as H$\rm \alpha$, often assume that the dominant contribution to observed emissions arises from star-forming activity, while the role of accreting black holes (BHs) remains largely unexplored. 
In this study, we use the IllustrisTNG cosmological hydrodynamical simulation to construct 
mock intensity maps of H$\rm \alpha$ and \ion{He}{II}, including contributions from both star formation and BH accretion. 
We show that the BH contribution to the mean intensity is significant, reaching $\sim40$--60 per cent for H$\rm \alpha$ and $\sim 60$--80 per cent for \ion{He}{II} around cosmic noon. 
Owing to the large luminosity weight of rare, bright sources, BH-powered emission dominates the shot-noise component of the power spectrum, and significantly boosts small-scale clustering amplitude, particularly for \ion{He}{II}.
We assess the implications for forthcoming LIM surveys and show that SPHEREx can probe the BH-influenced bright end of the H$\alpha$ VID at $z\lesssim4$, and a CDIM-like experiment can further access the BH-dominated regime of \ion{He}{II}. Our results demonstrate that accreting BHs represent an essential component of LIM signals,
which was previously underappreciated.  
We thus conclude that accurately modeling the BH contribution is crucial for a physically complete interpretation of future LIM observations.
\end{abstract}




\begingroup
\let\clearpage\relax
\endgroup
\newpage

\section{Introduction}
Line Intensity Mapping (LIM) has been proposed as a powerful technique for probing the large-scale structure and cosmic history of luminous matter by statistically measuring the integrated emission from unresolved galaxies \citep[see the review by][]{Kovetz2017, Bernal2022, Chang2026}. The earliest LIM studies primarily focused on the 21 cm line of neutral hydrogen as a tracer of large-scale structure \citep[e.g.][]{Scott1990, Chang2008, Chang2010, Masui2013, Switzer2013, Anderson2018}. More recently, it has been extended to several other atomic and molecular emission lines ranging from the microwave to the ultra-violet (UV) bands, such as Ly$\alpha$ \citep[e.g.][]{Silva2012, Pullen2014, Comaschi2016, Heneka2017, Croft2018, Mas-Ribas2020}, \ion{He}{II} \citep[e.g.][]{Visbal2015, Parsons2022}, H$\alpha$ \citep[e.g.][]{Fonseca2017, Gong2017, Silva2018, Gong2020}, CO \citep[e.g.][]{Righi2008, Gong2011, Lidz2011, Keating2016, Keating2020, Li2016, Chung2019, Breysse2022, Stutzer2024}, and [\ion{C}{II}] \citep[e.g.][]{Gong2012, Silva2015, Pullen2018, Yang2019, Yue2019, Chung2020, Sun2021}. Several initial LIM detections have already been reported, including 21 cm at $z \sim 0.8$ \citep{Masui2013}, [\ion{C}{II}] at $z \sim 2.5$ \citep{Pullen2018}, and Ly$\alpha$ at $z \sim 3$ \citep{Croft2016, Croft2018} through cross-correlations with galaxy or quasar surveys, as well as a CO auto-spectrum detection at $z \sim 3$ in the shot-noise regime \citep{Keating2016,Keating2020}. Motivated by these initial detections, a number of ongoing and upcoming surveys aim to extend these measurements to larger volumes and higher redshifts \citep[e.g.][]{Crites2014, Concerto2020, Crill2020, CCAT-Prime2023}.

In conjunction with these forthcoming surveys, theoretical modeling efforts will require further refinement. Apart from the 21 cm line, the emission lines targeted by LIM are generally associated with galaxies. For these lines, most existing models assume that the targeted emission lines originate from nebular emission powered by star formation. 
In contrast, the possible contribution from accreting black holes (BHs) has received comparatively little attention, despite their potential to modify LIM observables. 
Their spectra are typically harder than those of stellar populations, allowing them to efficiently produce high-ionization emission lines. Moreover, luminous BHs preferentially reside in massive halos, whose strong clustering may enhance their imprint on LIM observables.

The potential contribution from accreting BHs has been briefly discussed in a limited number of studies in the context of LIM. 
For example, \citet{Silva2018} modeled the contribution of the BH-powered emission through an ad hoc boost to the mean H$\alpha$ intensity, motivated by observations at $z\sim1-2$. 
This provides a useful phenomenological estimate, but the BH contributions at higher redshifts and for summary statistics beyond the mean intensity, such as the effective bias and the voxel intensity distribution, have not yet been fully explored. 
Another example is provided by \citet{Visbal2015}, who incorporated BH-powered \ion{He}{II} 1640 \AA~ emission in their line modeling. However, their study targeted very high redshifts ($z \gtrsim 10$), primarily focusing on the effects of Pop III star formation, and the BH contribution itself was not the main focus.
In addition, active galactic nuclei (AGN) have also been considered as possible cross-correlation counterparts to LIM, but such studies generally do not incorporate BH-powered emission into the line-intensity models themselves \citep[e.g.][]{Breysse2019}.

In this study, we extend the theoretical modelling of LIM by incorporating BH-powered emission. We focus in particular on H$\alpha$ and \ion{He}{II} lines. H$\alpha$ is one of the prime targets of upcoming LIM surveys,  owing to its relatively high luminosity compared with other optical and UV emission lines. 
\ion{He}{II} has been less widely explored in the LIM context \citep[but see e.g.][]{Visbal2015, Parsons2022}, but is expected to receive a substantial contribution from BHs.
We utilize the IllustrisTNG cosmological hydrodynamical simulation suite \citep{Nelson2019}. 
We compute the line luminosities for individual galaxies, separately evaluating the contributions from star formation and BH accretion, and then construct mock LIM observables, including the voxel intensity distribution (VID), auto-power spectra, and cross-power spectra, to quantify the impact of BH-powered emission on the statistics of LIM signals. 

This paper is organized as follows. In Section~\ref{sec:Method and Model}, we describe the IllustrisTNG galaxy catalog used in this work and present our modelling framework for emission lines, including both the star-forming and BH-powered components of H$\alpha$ and \ion{He}{II}. We also describe the procedure used to construct the mock LIM maps. Section~\ref{sec:Results} presents the main results, including the contribution of BH-powered emission to the mean intensity, the voxel intensity distribution (VID), and the auto- and cross-power spectra, together with an assessment of their detectability. In Section~\ref{sec:Discussion}, we discuss several additional physical and observational effects, including dust attenuation, IGM emission, and the impact of AGN torus obscuration. Finally, Section~\ref{sec:Conclusion} summarizes our conclusions and outlines prospects for future work.

\section{METHOD AND MODEL}
\label{sec:Method and Model}
In this section, we describe the 
modelling framework used to construct mock line-intensity maps. 
We first extract galaxy populations from the IllustrisTNG cosmological magnetohydrodynamical simulation and obtain the relevant physical properties of each galaxy (Section \ref{sec:Galaxy Data}). Using these quantities, we compute the luminosities of the recombination lines H$\alpha$ 6563 \AA~ and \ion{He}{II} 1640 \AA~ using the emissivities computed by \texttt{PyNeb} (Section \ref{sec:modeling of emission lines}). 

\subsection{Galaxy Data: IllustrisTNG}
\label{sec:Galaxy Data}
We use outputs from the IllustrisTNG cosmological magnetohydrodynamical simulation suite \citep{Marinacci2018, Naiman2018, Nelson2018, Pillepich2018b, Springel2018}. The simulations are performed with the moving-mesh code \texttt{AREPO} \citep{Springel2010, Pakmor2011, Pakmor2016}, which solves the equations of ideal magnetohydrodynamics for the gas component while evolving other components (dark matter, stars, and BHs) with an N-body scheme, coupled through Newtonian self-gravity. The adopted cosmological parameters are consistent with the \textit{Planck} results \citep{Ade2016}. 


Among the several simulation runs of TNG, we use the galaxy catalog from TNG300-1, the highest-resolution run of the largest simulation volume, which follows the formation and evolution of cosmic structure in a periodic box of side length $205\,h^{-1}\mathrm{cMpc}$ ($302.6\,\mathrm{cMpc}$).
The baryonic mass resolution of TNG300-1 is $1.1\times10^{7}\,M_{\odot}$ and the dark matter particle mass is $5.9\times10^{7}\,M_{\odot}$. This corresponds to a minimum resolved halo mass of order $\sim10^{9}M_{\odot}$. 

The IllustrisTNG simulations include a comprehensive set of subgrid models that capture the essential physics of galaxy formation and evolution, including gas thermodynamics in a UV background, stellar evolution with chemical enrichment, and the seeding and growth of massive BHs. These models have been calibrated by comparison with a range of observational constraints. In particular, they reproduce several galaxy scaling relations measured at $z = 0$; the galaxy stellar mass function, the halo-to-stellar mass relation, the total gas mass within the virial radius of massive groups, the stellar size–mass relation, and the galaxy–BH mass relation. In addition, the simulations reproduce the overall evolution of the cosmic star formation rate density up to $z \lesssim 10$ \citep[e.g.][]{ Pillepich2018a, Pillepich2018b}. 
We caution that the simulation has been primarily calibrated against observational constraints at $z = 0$, and our predictions for high-redshift galaxies, such as those in our present study, should be interpreted with appropriate care \citep[see e.g.][]{Weinberger2017}.

\subsection{Modeling of Emission Lines}
\label{sec:modeling of emission lines}
In this section, we describe how emission-line luminosities are assigned to galaxies identified in the IllustrisTNG simulation. For each galaxy, we separately evaluate the contributions from star formation and BH accretion (hereafter referred to as SF- and BH-powered emission, respectively). The luminosity of a given emission line $i$ for an individual galaxy is modelled as the sum of these two physically distinct components,
\begin{align}
    L_i = L_i^{\rm SF} + L_i^{\rm BH}.
\end{align}
The SF-powered component traces line emission from interstellar gas ionized by young stellar populations, while the BH-powered component traces line emission associated with gas ionized by radiation from BH accretion disc. The relative importance of these two components varies strongly across galaxy populations and cosmic time. 


The intrinsic line luminosity emitted by species $i$ ($i=$ H$\rm \alpha$ or \ion{He}{II}) can be expressed as
\begin{align}
    L_{i}
    =4\pi j_{i}V
    =\frac{4\pi j_{i}\dot{N}_{i}}{n_{\rm e}n_{i^+}\alpha_{\rm B,\it i}}~\rm erg~s^{-1},
    \label{eq:line_lumi}
\end{align}
where $4\pi j_i$ is the emission coefficient, 
$V$ is the emitting volume, 
$\dot{N}_i$ is the ionizing photon production rate, 
$n_{\rm e}$ is the electron number density, 
$n_{i^+}$ is the number density of the next higher ionization state, and $\alpha_{\rm B,\it i}$ is the case-B recombination coefficient. 
We evaluate the atomic physics quantities appearing in Equation~\ref{eq:line_lumi} with \texttt{PyNeb} \citep{Luridiana2013}. The effective emissivity $j_i/(n_{\rm e}n_{i^+})$ and the recombination coefficient $\alpha_{{\rm B},i}$ are calculated under typical nebular conditions of electron temperature $T_{\rm e}=10^4~\mathrm{K}$ and electron number density $n_{\rm e}=10^2~\mathrm{cm^{-3}}$. 
The recombination coefficients and emissivities depend only weakly on $T_{\rm e}$ and $n_{\rm e}$, and moderate deviations from these fiducial values do not significantly affect the inferred line luminosities.
\footnote{The case-B recombination coefficient approximately scales as $\alpha_{\rm B}\propto T_{\rm e}^{-0.8}$, and the emissivities of the H$\alpha$ and \ion{He}{II} recombination lines scale as $j_i/(n_{\rm e}n_{i^+})\propto T_{\rm e}^{-(0.8\text{--}1)}$.} 
In Section~\ref{sec:Stellar component} and \ref{sec:Black hole component},
we describe how the ionizing photon production rate $\dot{N}_i$ is estimated for the stellar and BH components.

It is worth noting that the \ion{He}{II} signal may be contaminated by stellar sources such as Population III (Pop III) stars, as discussed in several previous studies \citep[e.g.][]{Oh2001, Visbal2015, Parsons2022}. However, these contributions are expected to be important primarily at high redshifts ($z \gtrsim 10$) or to remain subdominant at lower redshifts. In this study, we focus on the redshift range $z\lesssim 6$, where BH-powered emission is expected to play a more significant role. We therefore do not further investigate stellar sources such as Pop III stars as potential contaminants.


In the following, we mainly focus on the intrinsic line luminosities. 
We will discuss the attenuation effects in Section~\ref{sec:Attenuation and IGM Emission}. 
We also neglect contamination from line interlopers and other sources and leave a detailed treatment of them to future work, but we discuss their expected level of impact in Appendix~\ref{sec:Line Contamination}.

\subsubsection{Stellar component}
\label{sec:Stellar component}
For the stellar component, 
massive ($M_\star \gtrsim 10\,M_\odot$) and short-lived ($t_\star \lesssim 20~\mathrm{Myr}$) stars, such as O and early B stars, dominate the ionizing photon budget, and therefore recombination-line luminosities trace the nearly instantaneous SFR 
\citep[e.g.][]{Kennicutt1994, Madau1998, Lee2009, McQuinn2015}. 
The relation between $\dot N_{\rm HI}^{\rm SF}$ and SFR depends on the assumed stellar initial mass function (IMF). 
Since the exact form of the stellar IMF still remains uncertain, particularly at high redshift, we consider both Salpeter and Chabrier IMFs in our analysis.\footnote{
The IllustrisTNG simulations adopt a Chabrier IMF in the stellar evolution and feedback models. 
As a different IMF could modify the simulated galaxy population through its impact on stellar feedback, a fully self-consistent treatment would, in principle, require rerunning the simulation. In this study, we assume that the resulting changes in the simulated galaxy population are subdominant for simplicity, and only focus on the IMF dependence of the ionizing photon production rate per unit SFR. 
}

For solar abundances and the Salpeter IMF ($0.1$--$100\,M_\odot$), the calibration between $\dot N_{\rm HI}^{\rm SF}$ and the SFR is given by \citet{Kennicutt1998} 
\begin{align}
\dot N_{\rm HI}^{\rm SF}
= 9.3 \times 10^{52}
\left(\frac{\mathrm{SFR}}{M_\odot~\mathrm{yr}^{-1}}\right)
~\mathrm{s}^{-1}.
\end{align}
When adopting the Chabrier IMF, we multiply the normalization on the right-hand side by a factor of 1.64 \citep{Madau2014}. Following previous studies \citep[e.g.][]{Behroozi2013}, we add a mean-preserving log-normal scatter of 0.2 dex to this relation, 
to account for the variability in the underlying populations, for example due to differences in their star formation histories. 
We note that the log-normal scatter is a convenient approximation but may not fully capture the true distribution of galaxy properties. In particular, different galaxy populations (e.g., star-forming and quenched systems) may follow distinct relations, potentially producing more complex or even bimodal distributions \citep{Behroozi2019}.

To compute the \ion{He}{II} emission lines, we use the hardness ratio of He$^+$ to H ionizing photons, $Q^{\rm SF}\equiv\dot{N}_{\rm HeII}^{\rm SF}/\dot{N}_{\rm HI}^{\rm SF}$. 
Stellar population synthesis models generally predict $Q^{\rm SF}\lesssim10^{-2}$ (e.g. Figure~5 of \citealt{Schaerer2003}), although the exact value depends on the assumed stellar population properties, including the upper-mass cutoff of the IMF, metallicity, stellar age, and the treatment of binary evolution. Observational estimates of the ionizing spectral hardness in local star-forming galaxies typically fall in the range $Q_{\rm HeII}/Q_{\rm HI} \sim 10^{-3}$–$10^{-2}$ (e.g. Figure~5 of \citealt{Eldridge2022}), with a maximum value of $\sim0.02$. 
In this study, we adopt a constant value of $Q^{\rm SF}=0.02$ for the stellar component. 
Adopting $Q^{\rm SF}=0.02$ therefore corresponds to an approximate upper limit for stellar populations, and this provides a conservative estimate of the BH contribution by maximizing the possible stellar contribution to the \ion{He}{II} emission.
The \ion{He}{II} ionizing photon production rate from star formation is then given by
\begin{align}
\dot N_{\rm HeII}^{\rm SF}
= 1.9 \times 10^{51}
\left(\frac{\mathrm{SFR}}{M_\odot~\mathrm{yr}^{-1}}\right)
~\mathrm{s}^{-1}.
\end{align}
The corresponding \ion{He}{II} recombination-line luminosity is typically $\lesssim10$ per cent of the H$\alpha$ luminosity.

\subsubsection{Black hole component}
\label{sec:Black hole component}
For BH-powered emission, ionizing photons originate from the accretion discs of BHs and are therefore set by the mass accretion rate. We compute the emission luminosity by converting the instantaneous BH accretion rate provided by IllustrisTNG simulations ($\dot M_{\rm BH}$) into a bolometric luminosity ($L_{\rm bol}^{\rm BH}$). We then deriving the corresponding ionizing photon production rate using an AGN spectral energy distribution (SED). 

The bolometric luminosity is calculated following the radiative-efficiency prescription implemented in IllustrisTNG:
\begin{align}
L_{\rm bol}^{\rm BH} =
\begin{cases}
\epsilon_{\rm r}\,\dot M_{\rm BH}\,c^2 & \dot M_{\rm BH} \ge 0.1\,\dot M_{\rm Edd}, \\[3pt]
\left(10\,\dot M_{\rm BH}/\dot M_{\rm Edd}\right)^2 \times 0.1\,L_{\rm Edd} & \dot M_{\rm BH} < 0.1\,\dot M_{\rm Edd},
\label{eq:lbol_bh}
\end{cases}
\end{align}
where $\epsilon_{\mathrm{r}}$ is the radiative efficiency, $c$ is the speed of light, $L_{\mathrm{Edd}}$ is the Eddington luminosity for electron scattering, and  $\dot{M}_{\mathrm{Edd}} \equiv L_{\mathrm{Edd}} / (\epsilon_{\mathrm{r}} c^2)$ is the Eddington accretion rate. This prescription captures the transition to a radiatively inefficient regime at low accretion rates, where the bolometric luminosity scales non-linearly with the accretion rate \citep{Churazov2005,Hirschmann2014}. For high accretion rates, we assume a constant radiative efficiency of $\epsilon_{\rm r}=0.2$.

To convert the bolometric luminosity into the hydrogen-ionizing photon production rate, $\dot N_{\rm HI}^{\rm BH}$, we assume a power–law ionizing continuum of the form $L_\nu \propto \nu^{-\alpha}$. Under this assumption, the hydrogen-ionizing photon production rate is given by
\begin{equation}
\dot N_{\rm HI}^{\rm BH}
= \int_{\nu_{\rm HI}}^\infty \frac{L_\nu}{h\nu}d\nu
= \frac{L_{\nu_0}}{h\alpha}
\left(\frac{\nu_0}{\nu_{\rm HI}}\right)^{\alpha},
\label{eq:NHI_general}
\end{equation}
where $h$ is Planck’s constant, $\nu_{\rm HI}$ is the hydrogen ionization threshold, and $L_{\nu_0}$ is the specific luminosity at an arbitrary reference frequency $\nu_0$
\footnote{
We note that, in Equation~\ref{eq:NHI_general}, the integral is formally taken over all frequencies above the hydrogen ionization threshold ($\nu_{\rm HI}$ to $\infty$). In practice, for $\alpha > 0$, the integral is dominated by photons near $\nu_{\rm HI}$, while higher-energy photons contribute negligibly due to the declining spectral energy distribution and the decreasing photoionization cross section ($\sigma \propto \nu^{-3}$).
}
. We adopt a fiducial value of $\alpha = 1.7$ for the ionizing continuum slope. This choice is motivated by observational constraints on the shape of quasar ionizing continua \citep{Lusso2015}. 

In Equation \ref{eq:NHI_general}, we choose $\nu_0$ to correspond to a rest-frame wavelength of 1450 \text{\AA}. The specific luminosity at 1450 \text{\AA} is obtained from the bolometric luminosity as \citep{Shen2020} 
\begin{equation}
L_{1450}^{\rm BH}
= \frac{L_{\rm bol}^{\rm BH}}
{{\rm BC}_{1450}(L_{\rm bol}^{\rm BH})},
\end{equation}
where ${\rm BC}_{1450}(L_{\rm bol}^{\rm BH})$ is the bolometric correction calibrated from quasar SEDs (see Equation 5 and Table 1 of \citealt{Shen2020}). 
The correction adopts the same EUV extrapolation as that described in the previous paragraph. To account for quasar-to-quasar variability, we include a mean-preserving log-normal scatter of 0.1 dex in the bolometric correction, consistent with the dispersion reported by \citet{Shen2020}. 

The ionizing photon production rate of \ion{He}{II}, $\dot N_{\rm HeII}^{\rm BH}$, is obtained in the same manner as for the stellar component, using the hardness ratio  $Q^{\rm BH}\equiv\dot{N}_{\rm HeII}^{\rm BH}/\dot{N}_{\rm HI}^{\rm BH}$. 
For a power–law ionizing continuum, this ratio is given by 
$Q^{\rm BH}= \left({h\nu_{\rm HI}}/{h\nu_{\rm HeII}}\right)^{\alpha}$ 
where $\nu_{\rm HI}$ and $\nu_{\rm HeII}$ correspond to the ionization thresholds at 13.6 and 54.4 eV, respectively. 
Adopting $\alpha = 1.7$ yields $Q^{\rm BH} \simeq 0.1$, reflecting the substantially harder ionizing spectrum of accreting BHs compared to stellar populations. 
The resulting \ion{He}{II} luminosity produced by BH accretion is $L_{\rm HeII}^{\rm BH} \approx 0.45\,L_{\rm H\alpha}^{\rm BH}$, approximately an order of magnitude larger \ion{He}{II}/H$\rm \alpha$ ratio 
than expected from normal stellar populations. This quantitative difference highlights the diagnostic power of \ion{He}{II} emission for identifying BH-dominated systems in our mock catalog.


\subsection{Constructing Mock Intensity Maps} 
\begin{figure}
 \includegraphics[width=\linewidth]{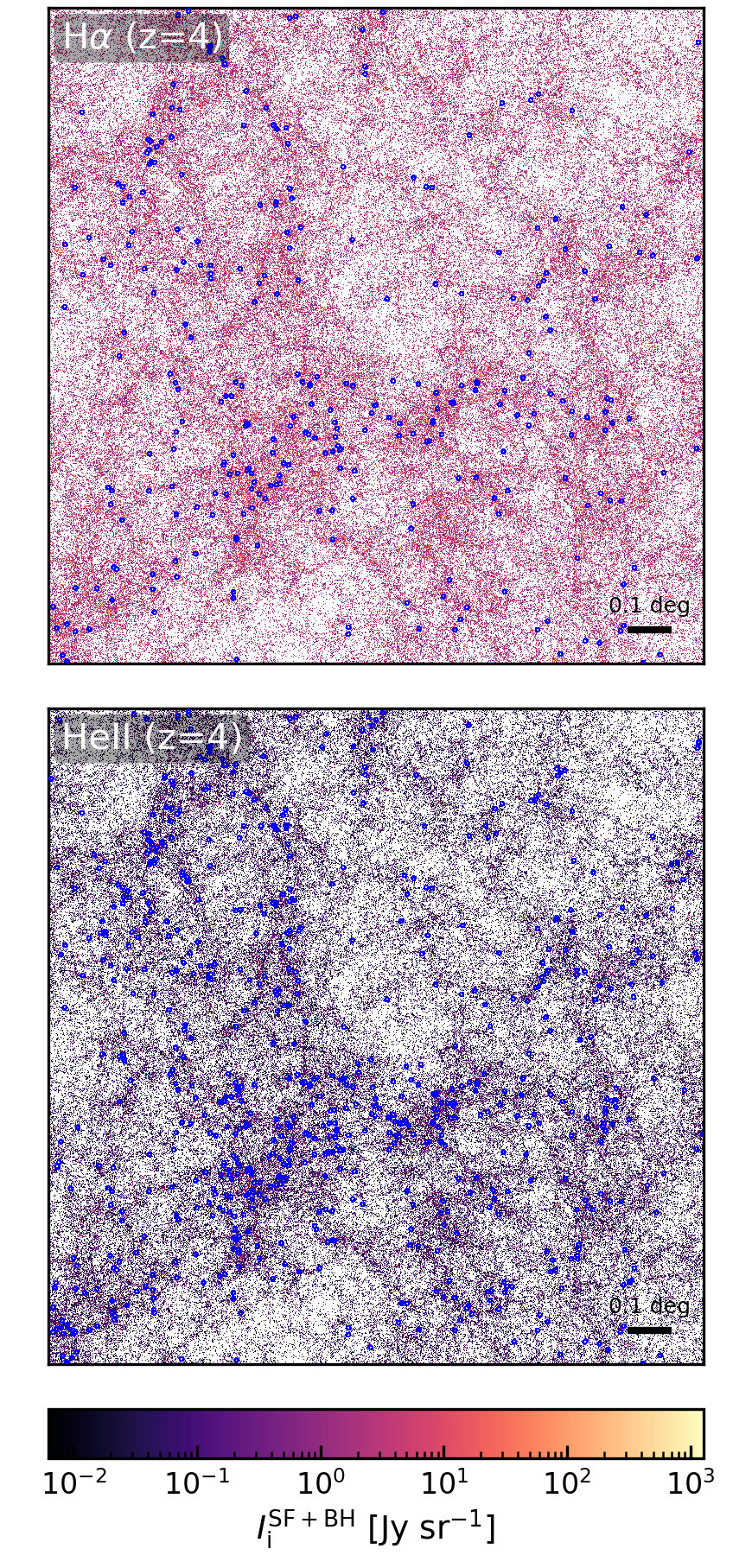}
 \caption{
 Intensity maps of H$\alpha$ (top) and \ion{He}{II} (bottom) emission at $z=4$ for a representative spectral bin in the SPHEREx-like mock survey. 
 The color scale shows the total intensity, including both SF- and BH-powered components. Blue points indicate regions where the BH-powered contribution dominates over the SF contribution within individual voxels.}
 \label{fig:SB_map}
\end{figure}
\begin{table}
	\centering
	\caption{
	Key specifications of the LIM experiments considered in this work.
	For the SPHEREx-like setup, we adopt the deep survey sensitivity,
	and for the CDIM-like setup we assume the ``improved''
	configuration following \protect\cite{Parsons2022}.
	}
	\label{tab:survey_specs}
	\begin{tabular}{lcc}
		\hline
		 & SPHEREx-like & CDIM-like\\
		\hline
		observed wavelength$^{\dagger}$ $[\mu m]$
        & $0.75 - 5$
        & $0.75 - 7.5$\\
		spectral resolution
        & $40$
        & $500$\\
        sensitivity $[\rm erg\,s^{-1}cm^{-2}Hz^{-1}sr^{-1}]$
        & $\sim 10^{-18}-10^{-19}$
        & $\sim 10^{-20}$\\
        survey area $[\rm deg^2]$
        & $200$
        & $31$\\
        pixel size $[\rm arcsec]$
        & $5$
        & $1$\\
        H$\alpha$ redshift range
        & $0.14-6.6$
        & $0.14-10.4$\\
        \ion{He}{II} redshift range
        & $3.6-29.5$
        & $3.6-44.7$\\
		\hline
	\end{tabular}

	\vspace{1mm}

	\raggedright
	\footnotesize
	$^{\dagger}$ To extend the \ion{He}{II} coverage down to
	$z\simeq1$, we additionally include a hypothetical auxiliary
	short-wavelength band spanning $0.25$--$0.93~\mu$m for both setups.
\end{table}

To create mock maps, we assign the line luminosity of each galaxy to observed-wavelength (redshift) bins. For the main analysis, we employ SPHEREx-like photometric bands  covering $0.75$–$5.0~\mu{\rm m}$. 
SPHEREx \citep[Spectro-Photometer for the History of the Universe, Epoch of Reionization, and Ices Explorer ;][]{Dore2014,Dore2018,Crill2024} is a NASA MIDEX mission launched on March 11, 2025, successfully completed the in-orbit checkout on May 1, 2025, and began its 25-month survey on the same day. 
The key instrumental parameters of SPHEREx are summarized in Table~\ref{tab:survey_specs}.
The actual SPHEREx instrument has a wavelength-dependent spectral resolution; however, for simplicity, we adopt a constant resolution $R \equiv \lambda / \Delta\lambda = 40$. 
To construct intensity maps, we divide the simulation volume into voxels with an angular size of $5'' \times 5''$ and a spectral width $\Delta \lambda$, and assign each galaxy's luminosity to the voxel containing the galaxy.


In addition to the SPHEREx-like setup, we also consider a proposed observational instrument, Cosmic Down Intensity Mapper \citep[CDIM;][]{Cooray2019}, which covers the wavelength range $0.75$–$7.5~\mu$m.
We construct mock maps adopting an angular resolution of $1''\times1''$ and a spectral resolution of $R=500$, following previous studies \citep[e.g.][]{Parsons2022}. The adopted survey specifications are summarized in Table~\ref{tab:survey_specs}.

To extend the \ion{He}{II} coverage down to $z\simeq1$, we additionally include an auxiliary short-wavelength band spanning $0.25$–$0.93~\mu$m for both the SPHEREx-like and CDIM-like setups. We assume the same spectral resolution as in each respective setup and treat this band as a hypothetical future UV–optical intensity-mapping observation. This auxiliary band is used only to extend the redshift coverage of \ion{He}{II} and to provide a complete view of its overall redshift evolution.


The resulting mock intensity maps serve as a physically motivated baseline for interpreting the fundamental behavior of line-intensity fluctuations and for quantifying how additional processes modify the observable signal. Validation tests and sanity checks of our numerical implementation, including consistency checks against observed luminosity functions, are presented in Appendix~\ref{sec:Sanity check}.

\section{Results}
\label{sec:Results}
\begin{figure}
 \includegraphics[width=\columnwidth]{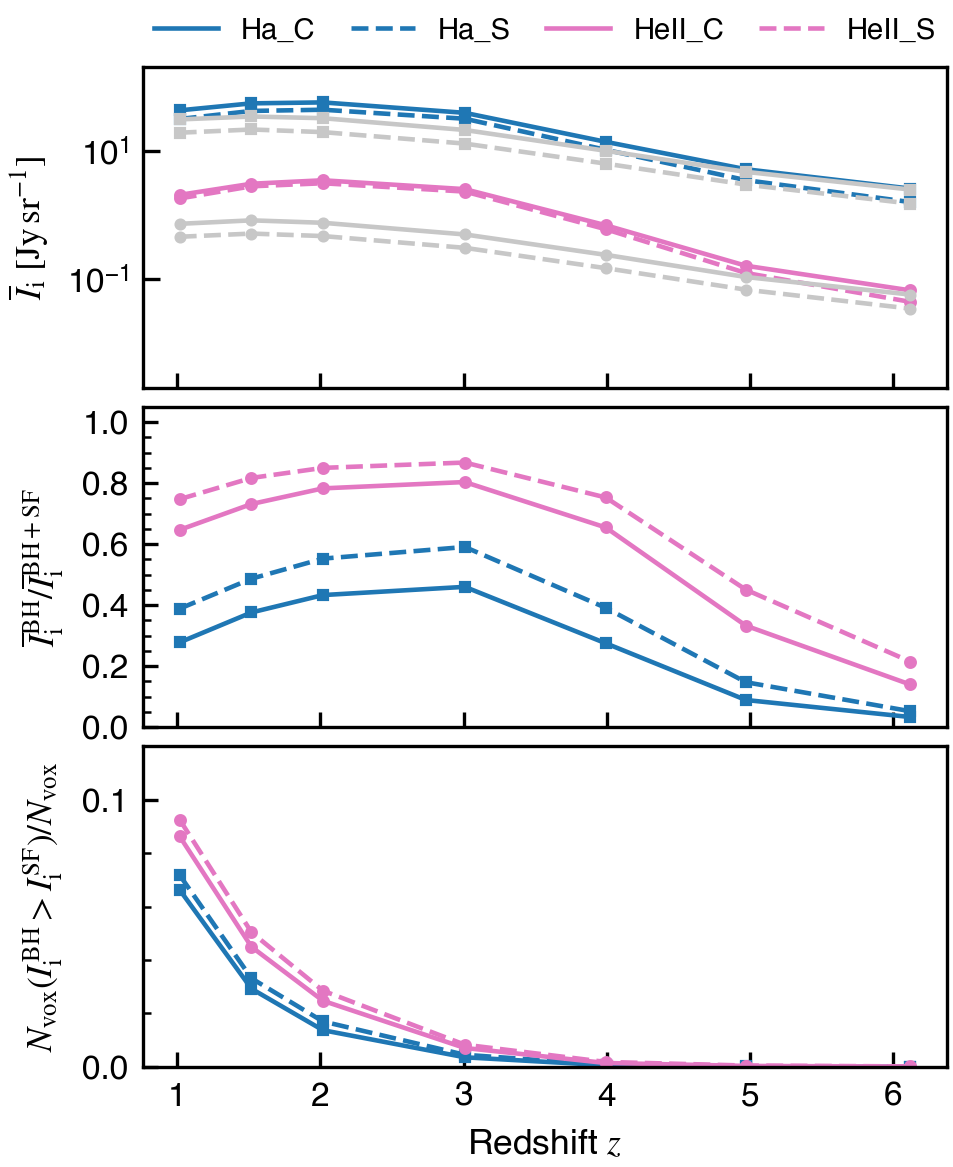}
 \caption{
 Redshift evolution of the mean line intensity and the role of BH-powered emission for H$\alpha$ and \ion{He}{II}. Top panel: mean intensity $\overline{I}_{i}$ for H$\alpha$ (blue) and \ion{He}{II} (pink). Colored curves show the total intensity (BH+SF), while grey curves indicate the SF-powered component; solid and dashed lines distinguish the Chabrier (H$\alpha$\_C and HeII\_C) and Salpeter (H$\alpha$\_S and HeII\_S) models, respectively. Middle panel: fractional contribution of BH-powered emission to the mean intensity, $\overline{I}_{i}^{\rm ~BH}/\overline{I}_{i}^{\rm ~BH+SF}$. Bottom panel: fraction of voxels in which the BH-powered brightness exceeds the SF contribution, defined as $N_{\rm vox}(I_{i}^{\rm BH}>I_{i}^{\rm SF})/N_{\rm vox}$.
 }
 \label{fig:meanI}
\end{figure}
\begin{figure*}
 \includegraphics[width=\textwidth]{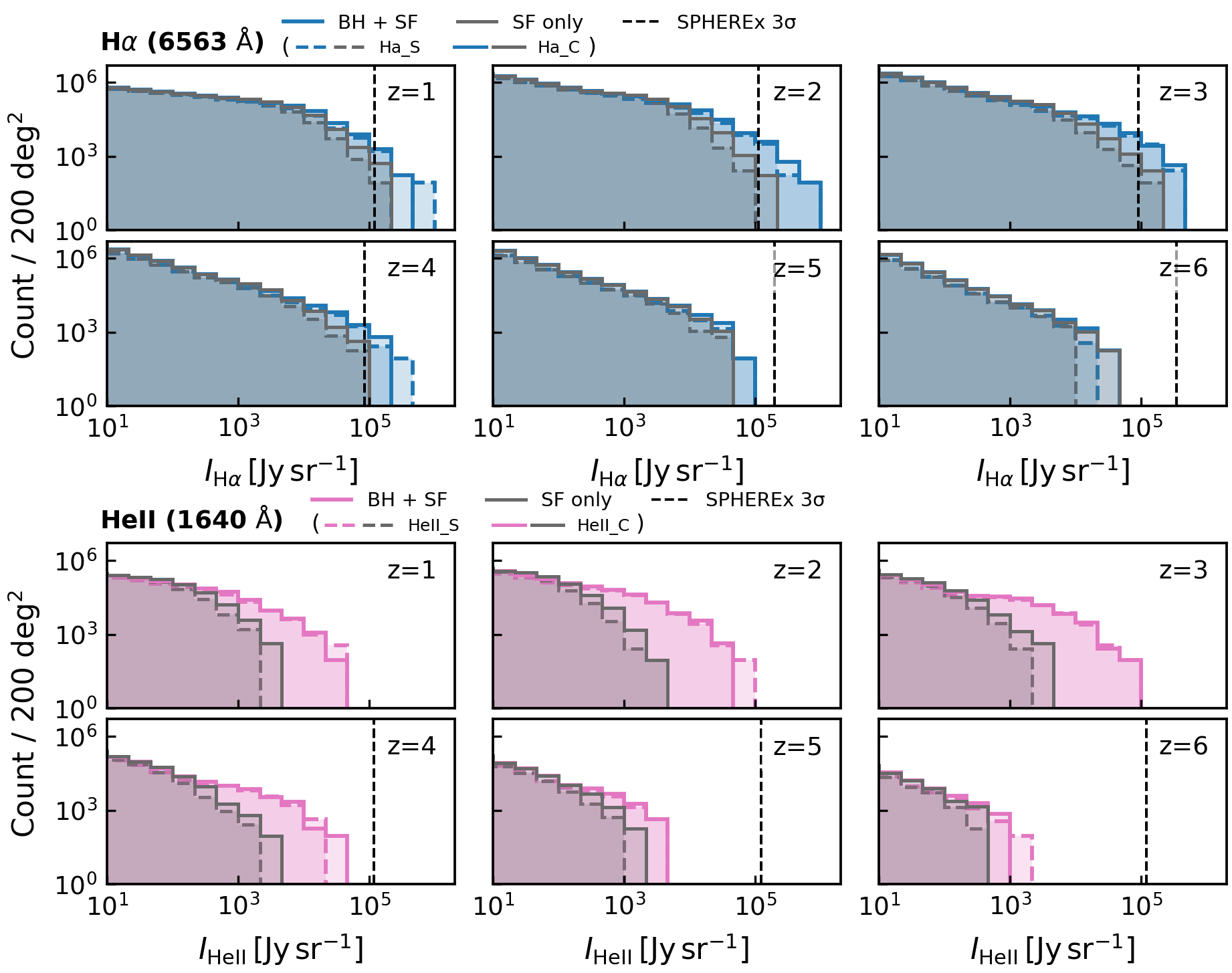}
 \caption{
 Voxel intensity distributions (VIDs) of the H$\alpha$ (top two rows) and \ion{He}{II} (bottom two rows) at $z=1-6$, shown for a representative spectral bin in each redshift slice. The distributions are rescaled to an effective survey area of 200 deg$^{2}$, corresponding to the SPHEREx deep field. Colored histograms show the total emission (BH+SF), while grey histograms indicate the contribution from SF alone. Solid and dashed outlines correspond to models assuming Chabrier and Salpeter IMFs, respectively. Vertical dashed lines denote the SPHEREx $3\sigma$ surface-brightness thresholds at the corresponding observed wavelengths.
}
 \label{fig:VID_count_Ha_HeII}
\end{figure*}
Figure~\ref{fig:SB_map} shows surface-brightness maps of H$\alpha$ and \ion{He}{II} emission at $z=4$, a representative redshift at which \ion{He}{II} becomes accessible to SPHEREx surveys. 
The line emission is computed assuming Chabrier IMF. 
Blue points in Figure~\ref{fig:SB_map} indicate regions where the BH-powered contribution dominates over the SF contribution within individual voxels.
One can see that BH-dominated voxels tend to coincide with overdense regions of the galaxy distribution. 
This result suggests that BH-powered emission can disproportionately enhance large-scale fluctuations in the intensity field.
In addition, as expected, the number of BH-dominated voxels is more numerous in \ion{He}{II} than in H$\alpha$, due to the much harder ionizing spectra associated with the accreting BHs. 

In the following sections, we quantify the impact of BH-powered emission on the statistical properties of the intensity maps. The results are labeled as Ha\_C and Ha\_S for H$\alpha$, and HeII\_C and HeII\_S for \ion{He}{II}, where the suffixes "C" and "S" denote Chabrier and Salpeter IMFs, respectively.

\subsection{Mean Intensity}
\label{sec:Mean Intensity}
Figure~\ref{fig:meanI} summarizes the redshift evolution of the mean line intensity and the relative importance of BH-powered emission for H$\alpha$ and \ion{He}{II}. The top panel shows the mean intensity, the middle panel presents the fractional contribution of BH-powered emission to the mean intensity, and the bottom panel shows the fraction of voxels in which the BH-powered intensity exceeds that of the SF-powered component.
Here, we focus on the intrinsic emission properties without including additional effects such as dust attenuation, ionizing-photon escape, or IGM recombination emission. These effects are discussed separately in Section~\ref{sec:Attenuation and IGM Emission}.

The top panel shows that the mean intensities of H$\alpha$ and \ion{He}{II} exhibit broadly similar redshift evolution, with a gradual rise from high redshift toward cosmic noon and a flattening or mild decline at lower redshift. 
The evolution of \ion{He}{II}, however, is somewhat steeper, reflecting its stronger dependence on BH-powered emission.
As seen in the middle panel, H$\alpha$ remains dominated by SF-powered emission, with the BH intensity fraction reaching at most $\sim40$ ($\sim 60$) per cent for a Chabrier (Salpeter) IMF, while \ion{He}{II} line maintains a systematically higher fraction of $60-80$ per cent around cosmic noon. 
Accounting for BH-powered emission is therefore a key step toward a physically complete description of LIM signals, particularly for lines sensitive to hard ionizing radiation such as \ion{He}{II}.


We note that the peak BH intensity fraction of $\sim40$ per cent in H$\alpha$ is higher than the AGN contamination levels commonly assumed when converting integrated H$\alpha$ luminosity functions into star-formation rate densities. Observational studies typically adopt AGN fractions of $\sim10$ per cent at $z\sim0.8$–1 and $\sim20$ per cent at $z\sim1.5$–2.2 \citep[e.g.][]{Garn2010,Sobral2013,Sobral2016}. This discrepancy may indicate that BH-powered H$\alpha$ emission from faint AGN contributes more significantly than currently inferred from observations, since the AGN number fraction in our galaxy catalog is broadly consistent with observational estimates (see Appendix~\ref{sec:AGN Fraction as a Function of Ha Luminosity}).

The bottom panel of Figure~\ref{fig:meanI} focuses on the spatial distribution of regions in which BH-powered emission dominates. For each redshift slice, we consider all spatial voxels that contain at least one galaxy and compute the fraction of those voxels for which the BH-powered emission exceeds the SF component. For both H$\alpha$ and \ion{He}{II}, despite the substantial contribution of BH accretion to the mean intensity, the BH-dominated voxel fractions remain below a few per cent across most redshifts, indicating that only a small fraction of the total volume is directly influenced by strong BH activity. 

The BH contribution to the mean intensity peaks around $z \sim 2-3$ (middle panel), whereas the fraction of BH-dominated voxels increases monotonically toward lower redshift (bottom panel).
This difference arises because the two statistics probe fundamentally different aspects of the BH contribution. The mean intensity is a luminosity-weighted quantity and is therefore strongly influenced by a small number of extremely luminous BHs, whose activity peaks at cosmic noon and subsequently declines toward lower redshift. 
In contrast, the number of BH-dominated voxels depends only on whether the BH-powered emission exceeds the SF-powered component on a voxel-by-voxel basis, and is therefore primarily sensitive to the relative strength of BH- and SF-powered emission, rather than the absolute luminosity of individual sources. 
Therefore, while the mean intensity mainly reflects the evolution of the most luminous BHs, the fraction of BH-dominated voxels primarily traces the more numerous galaxy population hosting moderately luminous BHs.


\subsection{Voxel Intensity Distribution}

We compute the voxel intensity distribution (VID) directly from our mock intensity maps. 
Figure~\ref{fig:VID_count_Ha_HeII} shows the voxel counts per a spectral slice, renormalized so that the counts correspond to an effective survey area of $200\, \mathrm{deg}^2$, matching the SPHEREx deep field.
Each panel compares the VIDs including BH-powered emission (colored histograms) with that arising from SF-powered emission alone (grey histograms). Differences between the assumed stellar IMFs (Salpeter and Chabrier) are indicated by different line styles.

For H$\alpha$, the SF component dominates the VIDs over most of the intensity range, while the BH-powered emission contributes primarily at the bright end. 
At low to intermediate redshifts ($z \sim 1-4$), the inclusion of BH-powered emission enhances the bright tail, while at $z \gtrsim 5$, the total (BH+SF) and SF-only VIDs become nearly indistinguishable. 
This reflects the fact that massive, rapidly accreting BHs are still rare at high redshifts, reducing the BH-powered contribution to the bright-end emission.
In addition, the assumed IMF further modulates the relative importance of BH-powered emission. As a Chabrier IMF produces higher SF-powered luminosities for fixed galaxy properties, the relative contribution of BH-powered emission is reduced, leading to both lower intensities and a smaller number fraction of BH-dominated voxels compared to the Salpeter IMF.

On the contrary to H$\rm \alpha$, for \ion{He}{II}, VIDs are dominated by BH-powered emission over a wide range of intensities and redshifts. At low to intermediate redshifts ($z \simeq 1-4$), the inclusion of BH-powered emission extends the bright tail to values several orders of magnitude higher than those reached by the SF-only VID. At higher redshifts ($z \gtrsim 5$), the SF-only and total (BH+SF) VIDs approach each other overall, reflecting the declining abundance and luminosity of actively accreting BHs. Nevertheless, BH-powered emission continues to dominate the highest-intensity voxels.


\label{sec:Detectability of VID}


In Figure~\ref{fig:VID_count_Ha_HeII}, 
we further show the expected $3\sigma$ sensitivity of the SPHEREx deep survey (black dashed line). 
For H$\alpha$, SPHEREx can probe the bright end of the VID at $z \lesssim 4$. 
We find that this observable regime corresponds to intensities at which BH-powered emission contributes more than 50 per cent both to the total intensity and to the voxel population. In contrast, the bright end of the \ion{He}{II} VID remains largely inaccessible to SPHEREx, owing to its substantially lower intrinsic surface brightness. 

In order to probe fainter regions of the VID, we also consider the proposed CDIM survey (see Appendix~\ref{sec:VIDs for CDIM} for the CDIM predictions).
For H$\alpha$, CDIM can access not only the BH-dominated bright end but also intermediate-intensity regimes where we find the emission is predominantly produced by star-forming galaxies. For \ion{He}{II}, the improved sensitivity of CDIM allows access to the high-intensity side of the VID, which remains largely inaccessible to SPHEREx. 
We find that the detectable voxels are still largely contributed by BH-powered emission, with contribution fractions close to unity.
This indicates that CDIM can directly probe BH-driven \ion{He}{II} emission through intensity mapping, providing a clean observational window that is effectively unattainable with current-generation LIM experiments.

Since the bright end of the VID is dominated by BH-powered emission, an interesting possibility is to use line-intensity maps to identify regions that are likely to host AGN. These candidate regions could then be followed up with targeted observations aimed at detecting individual AGN. As BH-dominated voxels tend to be located in high-intensity regions (see Figure~\ref{fig:SB_map}), incorporating spatial information in addition to individual voxel intensities may improve the identification of promising AGN candidates.

\subsection{Effective Bias}
\label{sec:Effective Bias}
\begin{figure}
 \includegraphics[width=\columnwidth]{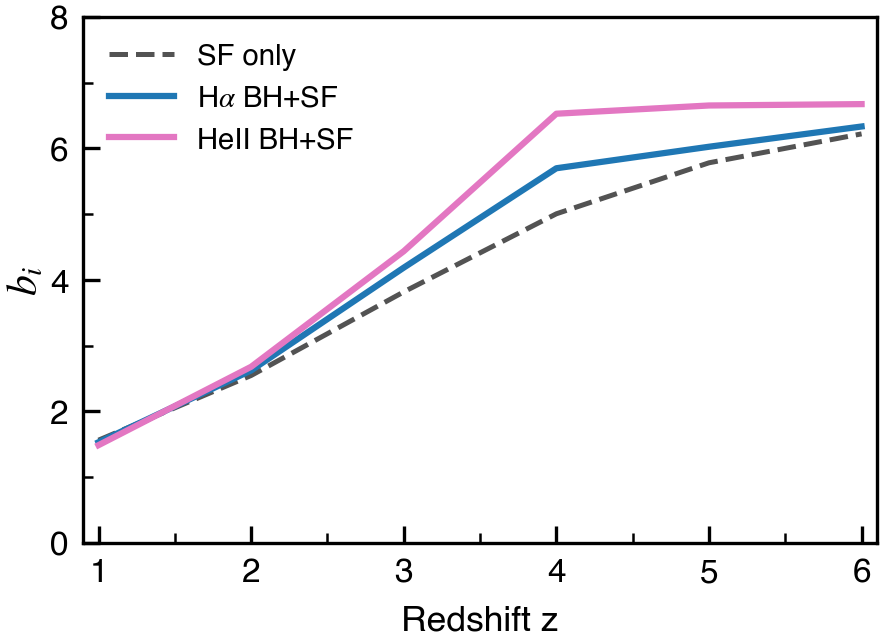}
\caption{
Redshift evolution of the effective bias, $b_i$, for H$\alpha$ and \ion{He}{II} emission in the Chabrier-IMF model. The grey dashed curve shows the bias associated with the SF-powered component, while the blue and pink solid curves show the total biases for H$\alpha$ and \ion{He}{II}, respectively, including both SF- and BH-powered emission. 
}
\label{fig:effective_bias}
\end{figure}
In addition to the one-point statistics discussed above, it is also important to investigate how BH-powered emission affects the intensity power spectrum.
On large scales, the power spectrum scales with the square of the product of the mean intensity and the effective bias. 
The effects of BH-powered emission in the mean intensity have been discussed in Section~\ref{sec:Mean Intensity}; in this section, we discuss the BH effects in bias. 

We estimate the effective bias from the cross-power spectrum between the intensity field and the matter density field. On large scales, we assume
\begin{align}
P_{I_i,m}(k,z)
=
\bar{I}_i(z)\, b_i(z)\, P_m(k,z),
\end{align}
where $P_{I_i,m}$ is the cross-power spectrum between the intensity field of line $i$ and the matter density field, $\bar{I}_i$ is the mean intensity of line $i$, and $P_m$ is the matter power spectrum. We therefore compute
\begin{align}
b_i(z)
=
\frac{P_{I_i,m}(k,z)}{\bar{I}_i(z)\, P_m(k,z)},
\end{align}
evaluated at $k = 0.1~h\,{\rm cMpc}^{-1}$.\footnote{
The range of accessible large-scale modes is limited by the adopted simulation volume. We therefore estimate the effective bias using the largest-scale mode available in our analysis. We confirm that the resulting values are broadly consistent with the luminosity-weighted bias estimated analytically from the halo distribution.
}


Figure~\ref{fig:effective_bias} shows the redshift evolution of the effective bias for H$\alpha$ and \ion{He}{II} emission assuming a Chabrier IMF. The grey dashed curve corresponds to the SF-powered component, while the colored solid curves include both SF- and BH-powered emission. We find that BH-powered emission systematically increases the effective bias relative to the SF-only case. This behavior arises because actively accreting BHs preferentially reside in relatively massive and highly biased halos compared to the broader population of SF-dominated emitters (see also Figure~\ref{fig:SB_map}). The enhancement is modest for H$\alpha$, but becomes more significant for \ion{He}{II} at $z \gtrsim 2$. Quantitatively, the total effective bias exceeds the SF-only value by $\simeq2-14$ per cent for H$\alpha$ and by $\simeq6-29$ per cent for \ion{He}{II}.

Together with the findings in Section~\ref{sec:Mean Intensity}, we can discuss the effects of BHs on the large-scale auto power spectrum. For instance, at $z \sim 4$, the H$\alpha$ boosts in the mean intensity and the effective bias, $\sim 1.5$ and $\sim 1.1$, imply a large-scale power boost of $\sim (1.5\times1.1)^2\sim2.7$, while the corresponding \ion{He}{II} boosts, $\sim 3.2$ and $\sim 1.3$, imply $(3.2\times1.3)^2 \sim 16.5$.
These estimates demonstrate that BH-powered line emission can substantially enhance the large-scale clustering signal, with the effect being particularly pronounced for \ion{He}{II}.



\subsection{Small-Scale Power Spectrum}
\label{sec:power_spectrum}
\begin{figure*}
 \includegraphics[width=\textwidth]{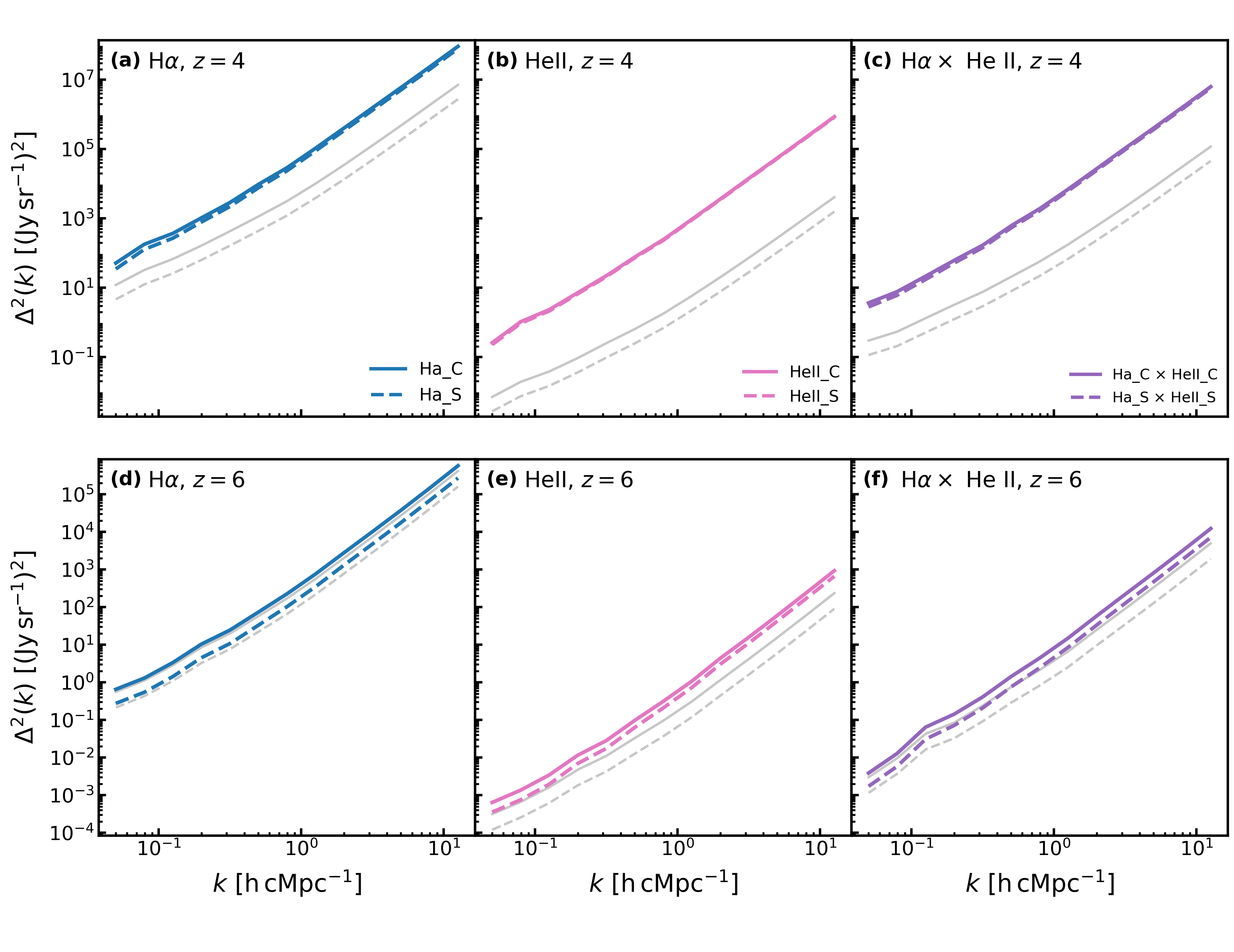}
 \caption{
 Auto- and cross-power spectra of the H$\alpha$ and \ion{He}{II} intensity fields at redshifts $z=4$ (top row) and $z=6$ (bottom row). Each row shows, from left to right, the auto-power spectra of H$\alpha$, \ion{He}{II}, and their cross-power spectrum. In all panels, the colored curves indicate the total emission including both BH-powered and SF-powered components, while grey curves show the contribution from SF alone.
}
 \label{fig:PS_Ha_HeII_z4_z6}
\end{figure*}
\begin{figure*}
 \includegraphics[width=\textwidth]{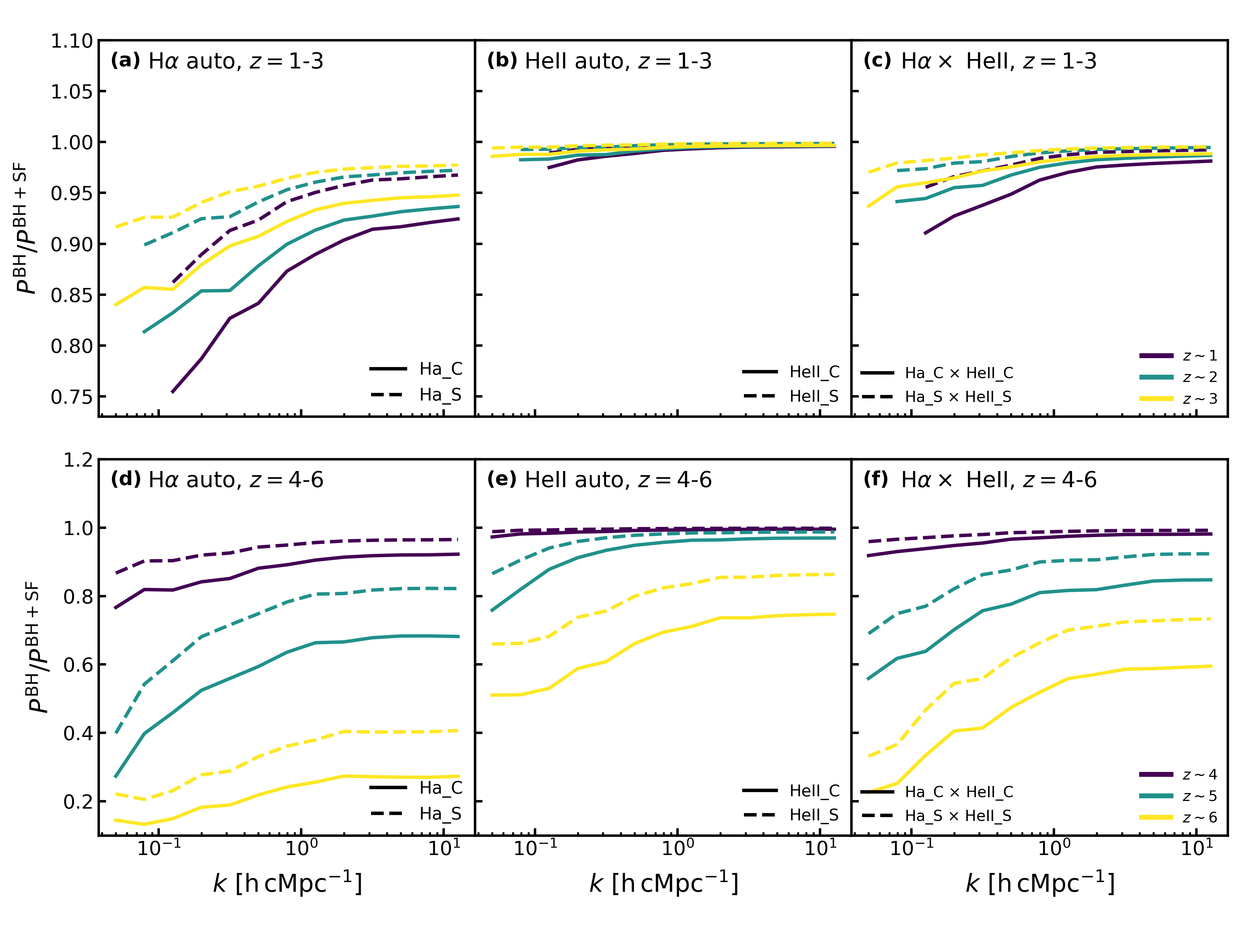}
 \caption{
 Scale dependence of the fractional contribution of BH-powered emission to the three-dimensional power spectra of the H$\alpha$ and \ion{He}{II} line-intensity fields. From left to right, the panels show the auto-power spectra of H$\alpha$, the auto-power spectra of \ion{He}{II}, and their cross-power spectra. The top row corresponds to the redshift shells centred at $z\simeq1-3$, while the bottom row shows the results for $z\simeq4-6$. In each panel, coloured curves represent different redshift, and different line styles distinguish the two stellar IMF models (C and S). The vertical axis shows the fractional contribution of BH-powered emission to the total power spectrum, where $P_{\rm BH+SF}$ and $P_{\rm SF}$ are the total and SF-only power spectra, respectively.
}
 \label{fig:PS_BHfraction}
\end{figure*}

We compute the three-dimensional power spectrum of each line-intensity field by performing a Fourier transform of the simulated three-dimensional intensity map $I_i(x, y, \nu)$. 
In addition to the auto-power spectrum, we also consider the cross-power spectrum between different emission lines, which can provide information on the underlying ionizing sources and help mitigate observational systematics (see Section~\ref{sec:Cross Power Spectrum} for more details).

For each emission line, we calculate both the auto-power spectrum,
\begin{equation}
P_{\rm auto}(\mathbf{k}) = \frac{\langle I_i(\mathbf{k}) I_i^\ast(\mathbf{k}) \rangle}{V} ,
\end{equation}
and the cross-power spectrum,
\begin{equation}
P_{\rm cross}(\mathbf{k}) = \frac{\langle I_i(\mathbf{k}) I_j^\ast(\mathbf{k}) \rangle}{V} .
\end{equation}
Here, $I_i(\mathbf{k})$ is the three-dimensional Fourier transform of the intensity field, $V$ is the volume of the analyzed region, and $\mathbf{k} = (k_\perp, k_{\parallel})$ denotes the wavevector, where $k_\parallel$ denotes the component of the wavevector along the line of sight and $k_\perp$ represents the transverse components on the sky. 
For simplicity, when computing the three-dimensional power spectrum, we choose the line-of-sight length of the computational domain to be comparable to the transverse length, to avoid an excessively anisotropic geometry in Fourier space (see Appendix~\ref{sec:Voxel Anisotropy}).
After obtaining the three-dimensional power spectrum $P(\mathbf{k})$, we bin the measured power in logarithmic intervals of $\Delta \log_{10} k = 0.2$ and average over spherical shells in $|\mathbf{k}|$ to derive the one-dimensional, angle-averaged power spectrum $P(k)$.

Figure \ref{fig:PS_Ha_HeII_z4_z6} summarizes the resulting angle-averaged power spectra, shown in terms of the dimensionless form
$\Delta^2(k) \equiv k^3 P(k)/(2\pi^2)$, for H$\alpha$ and \ion{He}{II} at $z=4$ (top row) and $z=6$ (bottom row). These redshifts are chosen because the wavelength coverage of SPHEREx allows the \ion{He}{II} line to be observed only at $z \gtrsim 4$. The left and middle columns show the auto-power spectra of each line and the right column presents the H$\alpha$–\ion{He}{II} cross-power spectra. In each panel, we compare the power spectra obtained from the total emission including both BH-powered and SF-powered components with those derived from the SF contribution alone. Figure~\ref{fig:PS_BHfraction} further quantifies the scale dependence of fractional contribution of BH-powered emission to the total power spectrum at $z = 1-3$ (top panels) and $z = 4-6$ (bottom panels). 

\subsubsection{Auto Power Spectrum}
We find that BH-powered emission enhances the amplitude of the power spectra for both H$\alpha$ and \ion{He}{II}, with the effect becoming particularly pronounced on small scales where the signal is dominated by shot noise from rare, luminous sources.

As seen in Figure~\ref{fig:PS_Ha_HeII_z4_z6}, the inclusion of BH-powered emission enhances the power spectrum in the shot-noise–dominated regime ($k \gtrsim 0.3-0.5\,h\,\mathrm{cMpc}^{-1}$ for H$\rm \alpha$ and $k \gtrsim 0.1-0.2\,h\,\mathrm{cMpc}^{-1}$ for \ion{He}{II})\footnote{
To quantify the scale at which the shot noise becomes dominant, we estimate the intrinsic shot-noise amplitude from the source population and compare it with the measured power spectrum. The shot-noise term is given by $P_{\rm shot} \propto \sum L^2$, and we define the transition scale by the condition $P_{\rm shot} \sim P_{\rm clust}(k)$. 
} 
by more than an order of magnitude for H$\alpha$, and by up to two orders of magnitude for \ion{He}{II}.
In Figure~\ref{fig:PS_BHfraction}, one can see that the BH contribution increases toward small scales in all cases.
The BH contribution in small-scale H$\rm \alpha$ auto-power is as large as 90 per cent at $z = 4$. The contribution rapidly decreases toward higher redshifts, but remains substantial even at $z=5$, with a value of $\gtrsim 70$ per cent.
For \ion{He}{II}, the contribution of BH to the small-scale power is always dominant, and it is as large as $\gtrsim 80$ per cent even at $z = 6$.

The large BH contribution at the shot-noise scales, a much higher contribution than one might expect from the modest BH contribution to the mean intensity, reflects the strong sensitivity of shot noise to rare, bright sources; the mean intensity scales as the first moment of the luminosity function, $\bar I \propto \sum_k L_k$, whereas the shot-noise term scales as the second moment, $P_{\rm shot} \propto \sum_k L_k^2$.
To further confirm this effect, we define the BH contributions to the mean intensity and to the shot-noise term as
\begin{equation}
f_{\bar I}=\frac{\sum_k L_k^{\rm BHdom}}{\sum_k L_k^{\rm BHdom}+\sum_k L_k^{\rm SFdom}},
\end{equation}
and
\begin{equation}
f_{\rm shot}=\frac{\sum_k (L_k^{\rm BHdom})^2}{\sum_k (L_k^{\rm BHdom})^2+\sum_k (L_k^{\rm SFdom})^2},
\end{equation}
where the sums run over individual sources within the survey volume. Here, $L_k^{\rm BHdom}$ and $L_k^{\rm SFdom}$ denote the line luminosities of sources belonging to the BH-dominated ($L^{\rm BH}>L^{\rm SF}$) and SF-dominated ($L^{\rm SF}>L^{\rm BH}$) populations, respectively.

Assuming that each population is characterized by a representative luminosity and number density, with $\eta \equiv n^{\rm BHdom}/n^{\rm SFdom}$, we obtain
\begin{equation}
f_{\rm shot}
=
\frac{f_{\bar I}^2/[(1-f_{\bar I})^2\eta]}
     {f_{\bar I}^2/[(1-f_{\bar I})^2\eta] + 1}.
\end{equation}
For representative values $f_{\bar I} \lesssim 0.4$ and $\eta \lesssim 0.06$, motivated by the H$\alpha$ results at $z=2$ (see Section~\ref{sec:Mean Intensity}), this relation yields $f_{\rm shot}\sim0.9$. This is in good agreement with the BH fractions shown in Figure~\ref{fig:PS_BHfraction}. This agreement indicates that the relatively modest BH contribution to the mean intensity, which is constrained by existing observations, is sufficient to produce a dominant shot-noise contribution

These results demonstrate that BH-powered emission constitutes a non-negligible component of the line-intensity power spectrum, particularly in the shot-noise–dominated regime. Since a primary science goal of line-intensity mapping is to probe the cosmic star-formation history, separating the contribution of BH-powered emission may be necessary for robustly inferring star-formation activity from LIM signals.

\subsubsection{Cross Power Spectrum}
\label{sec:Cross Power Spectrum}
As shown in Figure~\ref{fig:PS_Ha_HeII_z4_z6}, the inclusion of BH-powered emission systematically enhances the cross-power amplitude at both $z=4$ and $z=6$, in close analogy with the auto-power spectra. Figure~\ref{fig:PS_BHfraction} further shows that the BH contribution to the cross-power spectrum is generally large and increases toward smaller scales. At $z\simeq1-3$, the BH fraction is already high, rising from roughly $\sim90-95$ per cent on large scales to $\gtrsim98$ per cent on small scales. At higher redshift, the cross-power remains strongly affected by BH-powered emission: at $z\simeq4$, the BH contribution is still $\gtrsim90$ per cent; at $z\simeq5$, it is typically $\sim55-90$ per cent depending on scale and IMF model; and even at $z\simeq6$ it remains substantial, increasing from $\sim20-55$ per cent on large scales to $\sim60-75$ per cent on small scales.

Cross-correlation provides several key advantages. First, it offers practical advantages for mitigating observational systematics. Because instrumental noise and interloper contamination from foreground and background lines are largely uncorrelated between different tracers, their contribution is suppressed in the cross-power spectrum. As a result, the H$\alpha$--\ion{He}{II} cross-power spectrum provides a cleaner and more robust probe of the underlying correlated signal.

Second, combinations of auto- and cross-power spectra provide access to line-intensity ratios, which encode information about the underlying ionizing sources. 
On large scales, where both lines trace the same underlying density field, the power spectra can be approximated as $P_{ij}(k) \propto \bar I_i \bar I_j b_i  b_j P_m(k)$. Assuming that the effective bias factors of the two lines are similar, ratios such as $P_{ij}/P_{ii} \sim \bar I_j / \bar I_i$ provide an approximate estimate of the relative line intensities. 
As shown in Figure~\ref{fig:effective_bias}, the effective biases of H$\alpha$ and \ion{He}{II} are broadly similar at $z\lesssim2-3$, where this approximation is expected to be most reliable.
Because line intensities trace the production rate of ionizing photons, their ratios reflect the relative abundance of photons at different energies. In the present case, recombination lines such as H$\alpha$ and \ion{He}{II} are largely insensitive to gas metallicity, making the H$\alpha$-\ion{He}{II} ratio sensitive to the hardness of the ionizing radiation field, and thus providing a constraint on the relative contribution of BH-powered and SF-powered emission.

\subsubsection{Detectability of Power Spectrum}
\label{sec:Detectability for PS}
\begin{figure}
 \includegraphics[width=\columnwidth]{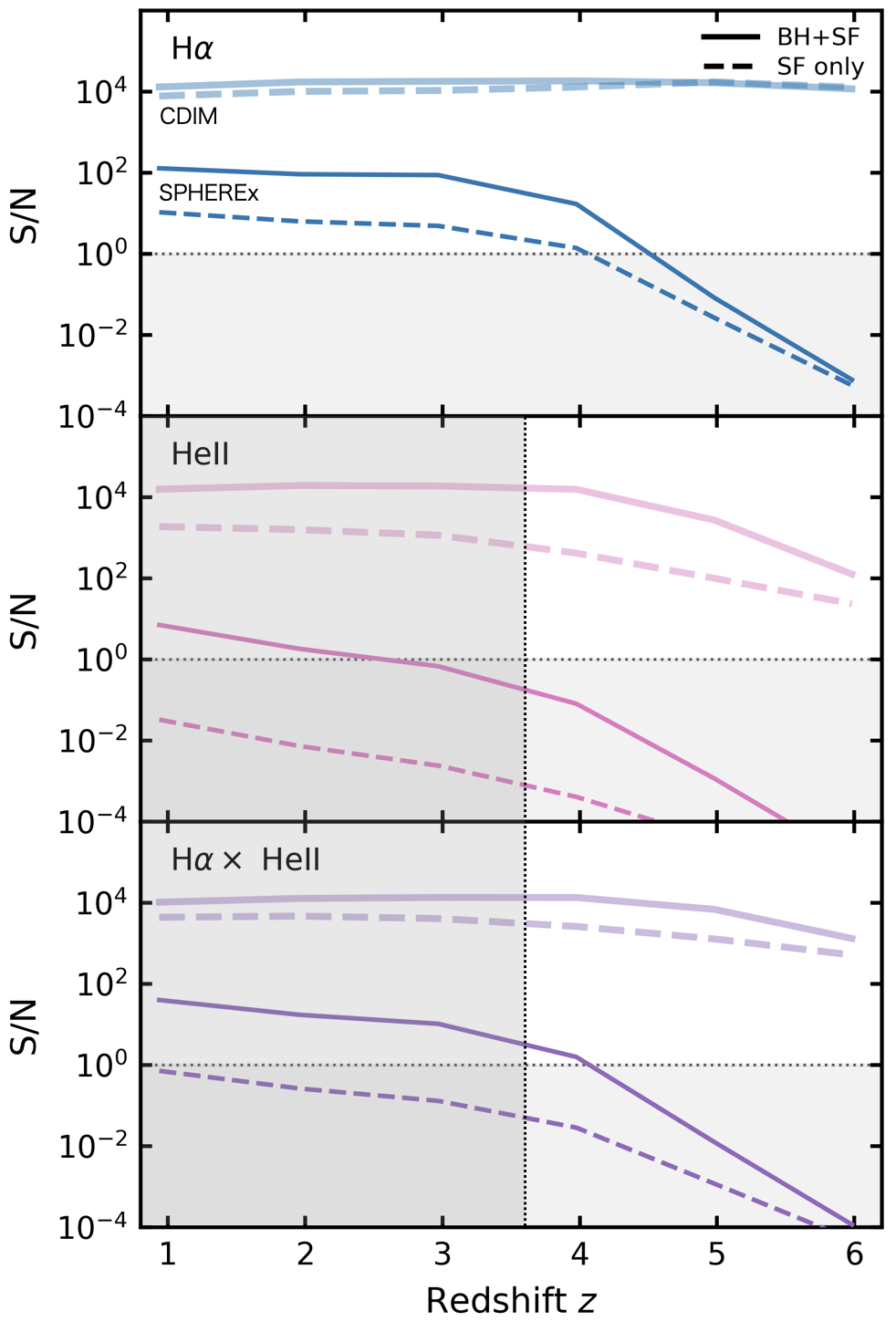}
\caption{
Integrated signal-to-noise ratio of the power spectra as a function of redshift for H$\alpha$ (top), \ion{He}{II} (middle), and the H$\alpha$–\ion{He}{II} cross-power spectrum (bottom), computed over the range $k_{\rm min}=0.1$ to $k_{\rm max}=10.0~h\,{\rm cMpc}^{-1}$. In each panel, the darker curves show the predictions for a SPHEREx-like survey, while the lighter curves show those for a CDIM-like survey. Solid and dashed curves correspond to the total (BH+SF) and SF-only models, respectively. The shaded horizontal region marks ${\rm S/N}<1$, and the shaded vertical region indicates the redshift range outside the survey coverage. All results are shown for the fiducial Chabrier-IMF model.
}
\label{fig:SN}
\end{figure}
We now assess the detectability of the H$\alpha$ and \ion{He}{II} power spectra for 
the SPHEREx- and CDIM-like experiments. 


For each $k$ bin, we estimate the statistical uncertainty of the power spectrum by combining the contributions from instrumental noise and sample variance. 
For the auto-power spectrum, we estimate the variance by neglecting the connected part of the four-point function of the intensity field, retaining only the disconnected Wick contractions.
\footnote{This approximation is expected to be most accurate on large scales, whereas connected-trispectrum contributions may become important in the shot-noise-dominated regime at small scales.}
Under this approximation, different Fourier modes are statistically independent, the covariance matrix becomes diagonal, and the variance in each $k$ bin is given by \citep[see e.g.][]{Sun2026}
\begin{equation}
\Delta P_{\rm auto}(k)
=
\frac{P_{\rm auto}(k) + P_{\rm N}/W^2(k,\mu)}{\sqrt{N_{\rm mode}(k)}},
\end{equation}
where $P_{\rm auto}(k)$ is the spherically averaged auto-power spectrum of the line-intensity field, $P_{\rm N}$ is the instrumental noise power spectrum, $W(k,\mu)$ is the window function describing the smoothing due to the finite angular and spectral resolution of the instrument, and $N_{\rm mode}(k)$ denotes the effective number of statistically independent Fourier modes contributing to the bin.

We assume that the instrumental noise is spatially uncorrelated, so that the corresponding noise power spectrum is white and can be written as
\begin{equation}
P_{\rm N} = \sigma_N^2\, V_{\rm vox},
\end{equation}
where $\sigma_N$ is the rms noise level per voxel in the intensity map and $V_{\rm vox}$ is the comoving volume of a single voxel. The value of $\sigma_N$ is determined from the instrumental sensitivity of each survey configuration considered in this work (SPHEREx or CDIM; see Section~\ref{sec:Detectability of VID}).

The factor $N_{\rm mode}(k)$ quantifies the number of statistically independent Fourier modes contributing to each $k$ bin. We evaluate $N_{\rm mode}(k)$ using the following analytic expression,
\begin{equation}
N_{\rm mode}(k)
=
\frac{k^2\,\Delta k\,V_{\rm survey}}{4\pi^2},
\label{eq:Nk}
\end{equation}
where $V_{\rm survey}$ is the comoving survey volume and $\Delta k$ is the width of the $k$ bin adopted in our power spectrum measurements. 

The window function $W(k,\mu)$ encodes the suppression of Fourier modes due to the finite angular and spectral resolution of the instrument. In LIM observations, this effect can be modeled as an anisotropic smoothing of the intensity field, with different characteristic scales in the transverse and line-of-sight directions. Assuming Gaussian instrumental responses, the window function can be written as
\begin{equation}
W(k,\mu)
=
\exp\left[
-\frac{k^2}{2}
\left(
(1-\mu^2)\sigma_\perp^2
+
\mu^2\sigma_\parallel^2
\right)
\right],
\end{equation}
where $\mu = k_\parallel/k$ is the cosine of the angle between the wavevector and the line of sight, and $\sigma_\perp$ and $\sigma_\parallel$ denote the transverse and line-of-sight smoothing scales, respectively. These are determined by the instrumental beam size and spectral resolution, and can be written as $\sigma_\perp = k_{\perp,\max}^{-1} \simeq \chi\,\theta_{\rm beam}$ and $\sigma_\parallel = k_{\parallel,\max}^{-1} \simeq (d\chi/d\nu)\,\Delta\nu$, where $\chi$ is the comoving radial distance, $\theta_{\rm beam}$ is the angular beam size, and $\Delta\nu$ is the spectral resolution.

For the cross-power spectrum between two intensity fields $i$ and $j$, we use the same definitions of $P_{\rm N}$ and $N_{\rm mode}(k)$. The corresponding uncertainty is estimated as
\begin{align}
\Delta P_{\rm cross}(k) 
&\approx \frac{1}{\sqrt{N_{\rm mode}(k)}}
\left[
  \frac{1}{2} P_{\rm cross}^2(k)
  + \frac{1}{2}
  \left( P_{i,{\rm auto}}(k) + \frac{P_{N,i}}{W_i^2} \right)
  \right. \notag\\
&\left.
  \times
  \left( P_{j,{\rm auto}}(k) + \frac{P_{N,j}}{W_j^2} \right)
\right]^{1/2}
\end{align}
where $P_{\rm cross}(k)$ is the cross-power spectrum, $P_{i,{\rm auto}}(k)$ and $P_{j,{\rm auto}}(k)$ are the auto-power spectra of the two fields, $P_{{\rm N},i}$ and $P_{{\rm N},j}$ are their respective instrumental noise power spectra, and $W_i(k,\mu)$ and $W_j(k,\mu)$ are the window functions. This expression is derived under the assumption that the Fourier modes of the intensity fields follow Gaussian statistics, as adopted for the auto-power spectrum. The first term inside the brackets represents the contribution arising from the intrinsic cross-correlation between the two intensity fields and contains no noise term, since the instrumental noises in the two maps are assumed to be uncorrelated. The second term corresponds to the contribution from the auto-correlations of the two fields and accounts for the variance induced by the total power in each field, including both the cosmological signal and instrumental noise.

Using these error estimates, we quantify the detectability of the power spectra in terms of the cumulative signal-to-noise ratio. For a given redshift, we define the total signal-to-noise ratio as
\begin{equation}
\rm S/N
=
\left[
\sum_{k \in [k_{\rm min},\,k_{\rm max}]}
\left(
\frac{P(k)}{\Delta P(k)}
\right)^2
\right]^{1/2},
\end{equation}
where the summation is taken over the $k$ bins in the range $k_{\rm min}=0.1$ to $k_{\rm max}=10~h\,{\rm cMpc}^{-1}$ for both the SPHEREx- and CDIM-like surveys. This adopted range lies within the accessible wavenumber range of the actual survey configurations.

Figure~\ref{fig:SN} shows the cumulative signal-to-noise ratio of the power spectra as a function of redshift. The H$\alpha$ auto-power spectrum is robustly detectable with high significance in a SPHEREx-like survey, reaching ${\rm S/N}\gtrsim10^2$ at $z\sim1$ and remaining well above unity up to $z\sim4$. Beyond this redshift, however, the signal-to-noise ratio drops rapidly, falling below the detection threshold at $z\gtrsim5$.
In contrast, the \ion{He}{II} auto-power spectrum is only marginally detectable at low redshift, with ${\rm S/N}\gtrsim10$ at $z\sim1$, and quickly becomes undetectable (${ \rm S/N}<1$) at $z\gtrsim2$. This reflects both the intrinsically lower intensity of \ion{He}{II} emission and the stronger impact of instrumental noise. We note \ion{He}{II} emission enters the observable wavelength range of SPHEREx only at $z\gtrsim4$.
The H$\alpha$–\ion{He}{II} cross-power spectrum exhibits an intermediate behavior, with ${\rm S/N}$ approaching unity at $z\sim4$ but remaining below the detection threshold over most of the redshift range. The cross-correlation is not affected by correlated instrumental noise, but its amplitude is insufficient to overcome the combined noise contributions from the two maps.

Figure~\ref{fig:SN} also compares the signal-to-noise ratios obtained from the total (BH+SF) and SF-only models. For H$\alpha$, the inclusion of BH-powered emission increases the detectability primarily at low redshift, although the overall behavior remains qualitatively similar between the two models. In contrast, the impact on \ion{He}{II} is much more pronounced: BH-powered emission boosts the signal-to-noise ratio by more than two orders of magnitude over a wide redshift range, making the difference between detectable and non-detectable signals in a SPHEREx-like survey. We note, however, that \ion{He}{II} enters the observable wavelength range of SPHEREx only at $z\gtrsim4$, so this improvement is not directly accessible to SPHEREx observations. A similarly strong enhancement is seen in the H$\alpha$--\ion{He}{II} cross-power spectrum, indicating that BH-powered emission plays a dominant role in determining the observability of these statistics.

A comparison with a CDIM-like survey demonstrates that improved sensitivity and spectral resolution significantly enhance the detectability of all statistics. In particular, while SPHEREx primarily enables a high-significance detection of the H$\alpha$ power spectrum, a CDIM-like survey is expected to extend detectability to \ion{He}{II} emission and the H$\alpha$–\ion{He}{II} cross-power spectrum over a broader redshift range.

Overall, these results indicate that SPHEREx will provide robust detections of the H$\alpha$ power spectrum, whereas probing \ion{He}{II} emission and cross-correlations, both of which are sensitive to BH-powered contributions, will require surveys with higher sensitivity such as CDIM.

\section{Discussion}
\label{sec:Discussion}

\subsection{Mitigating the AGN Contribution}
\label{sec:AGN_masking}
\begin{figure}
 \includegraphics[width=\columnwidth]{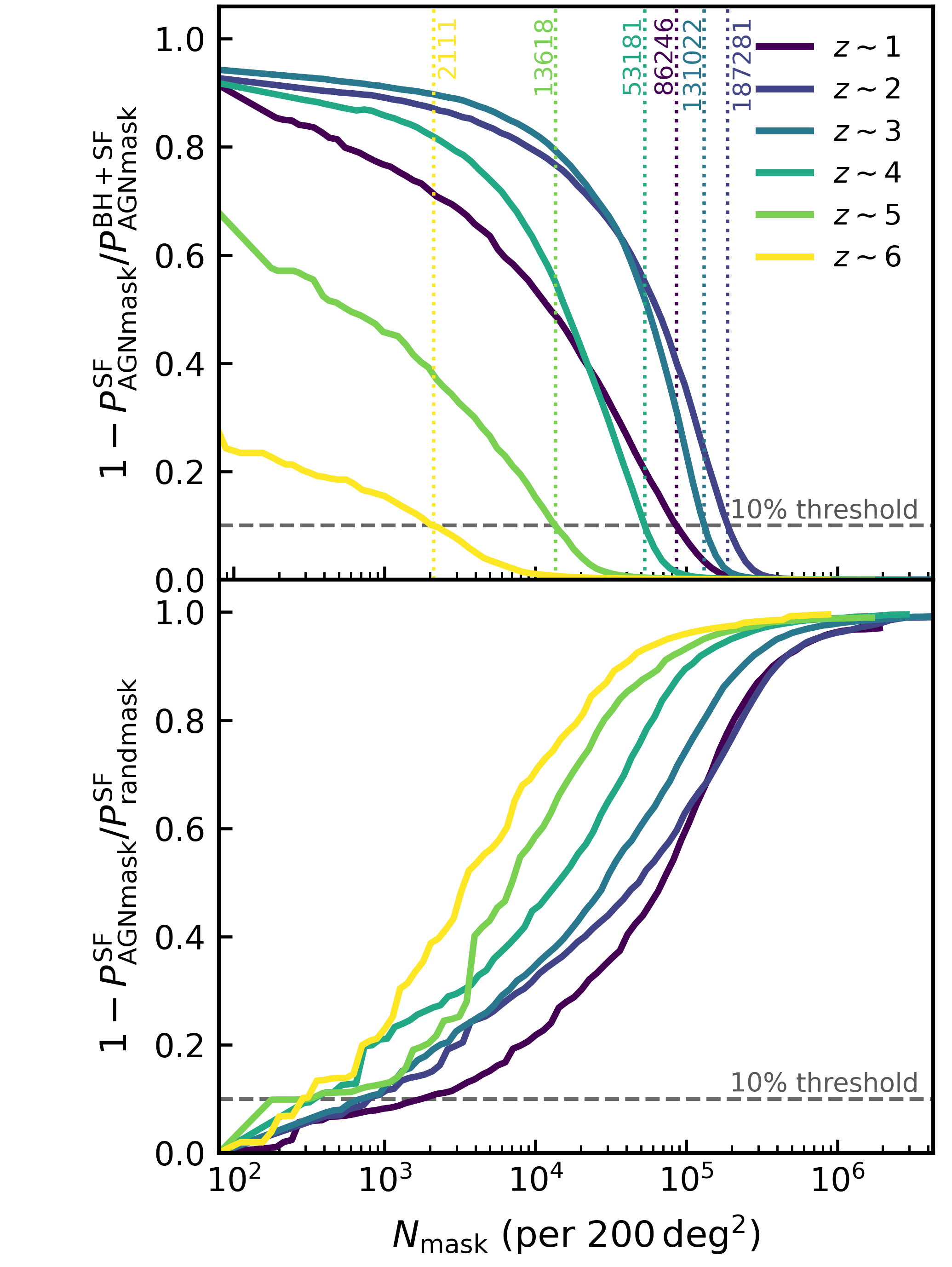}
\caption{
Impact of masking voxels associated with bright accreting BHs on the shot-noise component of the H$\alpha$ power spectrum. The horizontal axis shows the number of masked voxels per spectral slice, $N_{\rm mask}$, scaled to a SPHEREx-like survey area of $200,{\rm deg}^2$. Voxels are removed sequentially in order of decreasing BH bolometric luminosity.
Top panel shows the quantity $1 - P_{\rm AGNmask}^{\rm SF}/P_{\rm AGNmask}^{\rm BH+SF}$, which measures the residual contribution of BH-powered emission to the shot-noise amplitude of the masked total map. The horizontal dashed line indicates a residual BH contribution of $10$ per cent, and the vertical dotted lines mark the minimum number of masked voxels required to reduce the residual contribution below this level.
Bottom panel shows the quantity $1 - P_{\rm AGNmask}^{\rm SF}/P_{\rm randmask}^{\rm SF}$, where $P_{\rm randmask}^{\rm SF}$ is computed using random masks that remove the same number of voxels in each spectral slice as the AGN-targeted mask. This panel therefore quantifies the additional suppression of the SF-powered shot-noise component caused by preferentially masking BH-associated voxels, beyond the suppression expected from randomly removing the same number of voxels.
}
\label{fig:shotnoise_masking}
\end{figure}
The results obtained in our analyses demonstrate that BH-powered emission can significantly enhance the small-scale power spectrum, particularly in the shot-noise regime. 
This contribution carries valuable information about accreting BHs, but can also complicate the interpretation of LIM measurements when the primary goal is to trace the cosmic star-formation history.
It is therefore useful to explore whether the contribution from luminous BH-powered sources can be mitigated through simple observational strategies. In this subsection, we consider H$\rm \alpha$ only.

A common approach in LIM analyses is the masking of bright voxels associated with known galaxies or quasars identified in external catalogs \citep[e.g.][]{Sun2018, VanCuyck2023}.\footnote{Previous applications of this approach have primarily focused on mitigating line interloper contamination, rather than selectively suppressing the contribution from a specific source population.} 
This is typically implemented by applying a mask, or equivalently a window function, to the intensity map, thereby excluding contaminated voxels from the statistical analysis.
To mimic such an approach, we rank voxels according to the bolometric luminosity of the BHs they contain and progressively remove the brightest BH-hosting voxels from the map. In practice, we implement masking by setting the intensity of the selected voxels to zero. 
We measure the power spectrum of the masked map and denote it as $P^{\rm BH+SF}_{\rm AGNmask}$. We apply the identical mask to a map constructed from the SF contribution alone and measure the corresponding power spectrum, $P^{\rm SF}_{\rm AGNmask}$. 


Figure~\ref{fig:shotnoise_masking} summarizes the effect of this masking procedure on the shot-noise component of the H$\alpha$ power spectrum, which we define as the median power over modes with $k \ge 0.7\,{\rm Mpc}^{-1}$. 
The horizontal axis indicates the number of masked voxels per spectral slice. 
In top panel, the vertical axis shows $1 - P_{\rm AGNmask}^{\rm SF}/P_{\rm AGNmask}^{\rm BH+SF}$, 
which measures the residual contribution from BH-powered emission to the masked map. 
As the number of removed voxels increases, the residual BH contribution decreases and approaches zero. The horizontal dashed line marks a residual BH contribution of $10$ per cent, and 
the vertical dotted lines indicate the minimum number of voxels that must be masked in each spectral slice for the residual contamination to fall below this level.

Our masking implementation inevitably removes not only the BH-powered emission but also any co-spatial SF-powered contribution. It is therefore important to quantify the extent to which this procedure suppresses the shot-noise signal of the SF-powered component.
We compute the {\it true} power spectrum of SF-only map by applying a random mask to the SF-only map with the same number of masked pixels as the original masked map, and denote it as $P^{\rm SF}_{\rm randmask}$. 
This random mask accounts for the loss of modes due to masking, while isolating the additional effect of the mask being correlated with the SF-only emission, namely the selective removal of SF-powered signal.
The bottom panel of Figure~\ref{fig:shotnoise_masking} shows $1 - P_{\rm AGNmask}^{\rm SF}/P_{\rm randmask}^{\rm SF}$.
This quantity measures how masking suppresses the shot-noise amplitude of the SF-only map, which is the relevant signal component when the goal is to trace the cosmic star-formation history.
Notably, even in the absence of BH-powered emission, the shot-noise amplitude decreases as progressively more voxels are masked. This reduction arises because the masking removes SF-powered emission contained in the same voxels as the masked BH sources and reduces the effective area used to measure the power spectrum. Quantitatively, the suppression becomes substantial once $N_{\rm mask}\gtrsim10^{3}$, for which $1 - P_{\rm AGNmask}^{\rm SF}/P_{\rm randmask}^{\rm SF}$ reaches $\sim0.1$ at $z=1$ and increases toward the higher redshift. For higher masking levels, $N_{\rm mask}\sim10^{5}$, the suppression becomes severe, approaching unity depending on redshift. This demonstrates that the masking procedure itself introduces a non-negligible bias in the measured power spectrum, which must be taken into account when interpreting the masked measurements.

These results suggest that, although masking bright BH-hosting voxels can effectively reduce the BH contribution to the shot-noise component, it simultaneously removes a non-negligible fraction of the SF-powered emission. Therefore, the shot-noise amplitude measured from the masked map, in which the intensities of selected voxels are set to zero, does not directly correspond to that produced by the full population of star-forming galaxies. 

A more refined approach would be to remove only the BH-powered emission component within each voxel, rather than masking the entire intensity in the voxel. Implementing such a strategy in practice would require external AGN catalogs covering areas comparable to SPHEREx or CDIM. 
Current multi-wavelength surveys, including wide-area X-ray, infrared, and optical surveys, already detect large numbers of AGN, but their sky coverage and data characteristics remain heterogeneous.
For example, eROSITA (extended ROentgen Survey with an Imaging Telescope Array; \citealt{Merloni2012}), an X-ray all-sky survey, has identified a large population of AGN \citep[e.g.][]{Brunner2022, Liu2022, Salvato2022}. Wide-area optical surveys such as SDSS (Sloan Digital Sky Survey; \citealt{York2000}) and DESI (Dark Energy Spectroscopic Instrument; \citealt{DESI2016}) cover approximately $\sim0.25$ and $\sim0.3$ of the sky, respectively, and provide spectroscopic quasar catalogs \citep[e.g.][]{Lyke2020, Wu2022, Chaussidon2023}. Infrared surveys with WISE (Wide-field Infrared Survey Explorer; \citealt{Wright2010}) also span the full sky and identify large AGN samples based on color selection, although with limited redshift precision \citep{Assef2017}. These surveys differ significantly in their selection methods, spatial completeness, and redshift accuracy. Consequently, no single dataset simultaneously provides uniform AGN identification over LIM-scale areas with reliable three-dimensional information. Additional approaches may therefore be required, such as constructing AGN maps using machine-learning techniques trained on large-scale simulations and multi-wavelength survey data.

\subsection{Attenuation and IGM Emission}
\label{sec:Attenuation and IGM Emission}
In this section, we consider several physical processes that may alter the observed LIM signals, including dust attenuation, escape of ionizing photons, obscuration by the AGN dust torus, and recombination emission from the IGM. We first introduce simplified prescriptions for these processes in Section~\ref{sec:Modelling Attenuation and IGM Emission}, and then discuss their combined impact on the mean intensity and power spectrum in Section~\ref{sec:LIM Observables with Attenuated and IGM Modell}.

\subsubsection{Modelling Attenuation and IGM Emission}
\label{sec:Modelling Attenuation and IGM Emission}

\paragraph*{Attenuated Galaxy Emission.}
For the emission lines considered here, the observed luminosity of line-$i$ is written as
\begin{align}
L_{i,\mathrm{obs}}
=
f_{\rm ion}^{\rm dust}
(1-f_{\rm ion}^{\rm esc})
f_i^{\rm dust}
L_i,
\label{eq:attenuation}
\end{align}
where $f_{\rm ion}^{\rm esc}$ denotes the fraction of ionizing photons escaping into the IGM, while $f_{\rm ion}^{\rm dust}$ denotes the fraction of the remaining ionizing photons that survive dust absorption and ionize the ISM. The factor $f_i^{\rm dust}$ accounts for the dust attenuation of line-$i$ photons, and $L_i$ is the intrinsic line luminosity discussed above. Equation \ref{eq:attenuation} describes a model where dust- and ISM-free cavities around star-forming regions allow some ionizing photons to escape, whereas the rest are either absorbed by dust or consumed in ionizing the ISM.

Accurate modelling of $f_{\rm ion}^{\rm dust}$, $f_{\rm ion}^{\rm esc}$ and $f_i^{\rm dust}$ is challenging, as these quantities depend on the multi-scale structure of the ISM and are only weakly constrained by observations. We therefore adopt simple prescriptions. For $f_{\rm ion}^{\rm dust}$, we adopt a constant value of $0.5$ for all galaxies following \citet{Inoue2001}. This corresponds to the assumption that a half of the non-escaping ionizing photons contribute to nebular emission within H\,\textsc{ii} regions, while the remainder are absorbed by dust.
For $f_{\rm ion}^{\rm esc}$, we assign different values to non-AGN and AGN populations, where AGN are defined as galaxies with Eddington ratios $\lambda_{\rm Edd} \ge 0.1$. For non-AGN galaxies, we adopt a fixed value of $f_{\rm ion}^{\rm esc,nonAGN}=0.1$, while for AGN, we explore $f_{\rm ion}^{\rm esc,AGN}=0.1$, $0.5$, and $0.9$.\footnote{
In reality, the escape fraction and dust absorption of ionizing photons are expected to depend on photon energy and may therefore differ between \ion{H}{I}- and \ion{He}{II}-ionizing photons. For simplicity, however, we assume the same values for both in this study.
}
This setup is motivated by observational studies suggesting that typical star-forming galaxies exhibit relatively low escape fractions, whereas galaxies hosting AGN can show systematically higher values \citep{Mostardi2013,Grazian2018}. This contrast may reflect the impact of AGN-driven feedback, which can clear gas from the central regions and facilitate the escape of ionizing photons \citep[e.g.][]{Menci2008,Giallongo2012,Dashyan2018,Penny2018,Menci2019}.
For $f_i^{\rm dust}$, we adopt
\[
f_i^{\rm dust}=10^{-E_i/2.5},
\]
where $E_i$ denotes the attenuation magnitude for line $i$. We adopt $E_{\rm H\alpha}=1$ mag and $E_{\rm HeII}=3$ mag based on observational constraints \citep{Sobral2013,Gonzlez2025}. 

We assume that both the escape fraction and dust attenuation are independent of the redshift. In the case of dust attenuation, where our fiducial values are motivated by low-redshift observations, this assumption is likely conservative, since the dust content of galaxies is expected to decrease toward higher redshift. Thus, our model may underestimate the observed mean intensities at high redshift. 
For the escape fraction of ionizing photons, while the fiducial value adopted above is supported by reionization studies \citep[e.g.][]{Bouwens2015}, the typical observed values at low redshifts are often below this value \citep[e.g.][]{Flury2022}. If the typical escape fraction is smaller than $0.1$, the line luminosities become slightly larger. 

\paragraph*{Covering Factor of the AGN Dust Torus.}
In the standard unified model of AGN \citep[see reviews by][]{Antonucci1993, Netzer2015}, the central engine is surrounded by a geometrically and optically thick dusty structure, commonly referred to as the AGN torus. This axisymmetric structure possesses a sufficiently high column density to obscure the central source along certain lines of sight, giving rise to the classical distinction between Type~1 and Type~2 AGN \citep{Antonucci1985}. The obscuration is commonly parameterized by a covering factor, $f_{\rm C}$, which encodes the fraction of BH-powered emission that is obscured by the dusty torus. 

The covering factor has been estimated using various methods, such as the ratio of infrared dust emission to bolometric luminosity \citep{Maiolino2007, Treister2008, Mor2011, Mor2012, Roseboom2013, Lusso2013, Toba2021}, the equivalent width of the Fe~K$\alpha$ line \citep{Ikeda2009, Yaqoob2010, Brightman2011, Ricci2013}, the strength of the scattered X-ray continuum \citep{Ueda2007, Brightman2012}, and the number ratio of Type~I to Type~II AGN \citep{Ueda2003, Gilli2007, Hasinger2008, Brightman2012, Malizia2012}. These studies indicate that the torus covering factor exhibits a broad range of values, typically $f_{\rm C} \sim 0.1-0.5$. 

Here, we adopt a covering factor of $f_{\rm C}=0.5$, near the upper range of observational estimates, as a conservative choice. The obscuration model is applied only to active AGN, which we identify as BHs with an Eddington ratio $\lambda_{\rm Edd} \geq 0.1$. 
In practice, we implement this effect using a simple stochastic prescription: for each AGN, we randomly classify the source as obscured or unobscured according to the covering factor $f_{\rm C}$. In obscured systems, we reduce the BH-powered line luminosity by a factor $(1-f_{\rm BLR})$, assuming that only the emission originating from compact inner regions (broad line region) is strongly attenuated, while a more extended component (narrow line region) remains visible:
\begin{equation}
L_{i,\rm obs}^{\rm BH} =
\begin{cases}
L_i^{\rm BH}, & \text{with probability } 1-f_{\rm C},\\
(1-f_{\rm BLR})\,L_i^{\rm BH}, & \text{with probability } f_{\rm C}.
\end{cases}
\end{equation}
We adopt a fiducial value of $f_{\rm BLR}=0.8$, motivated by observations of the relative strengths of narrow and broad H$\alpha$ emission \citep{Stern2012}.


\paragraph*{IGM Emission.}
At the redshifts considered in this work ($z\leq6$), the hydrogen in IGM is highly ionized by the UV background, 
as indicated by various probes such as Ly$\alpha$ forest measurements \citep[e.g.][]{Dijkstra2014}, CMB Thomson scattering \citep[e.g.][]{Aghanim2018}, and absorption signatures in quasar and gamma-ray burst spectra \citep[e.g.][]{Ciardi2008}. Nevertheless, the mean free path of ionizing photons remains sufficiently short at these redshifts. For example, at the Lyman limit, it can be approximated as $\approx 85[(1+z)/4]^{-4}$ proper Mpc (e.g. \citealt{FaucherGigure2008, Gao2025}). This implies that most photons escaping from galaxies are eventually absorbed  in the surrounding medium.

To provide a simple estimate of the mean intensity, we consider a photon-counting argument, rather than modelling the detailed thermodynamic state of the IGM. For simplicity, we assume photoionization equilibrium, Case-B recombination, and neglect collisional excitation. Under these assumptions, the mean intensity reflects the number of ionizing photons absorbed by the gas. The mean intensity contributions from galaxies and the IGM then scale approximately as $\overline{I}_{i,\rm gal}\propto f_{\rm ion}^{\rm dust} (1-f_{\rm ion}^{\rm esc}) f_i^{\rm dust} \dot{N}_i$, $\overline{I}_{i,\rm IGM}\propto f_{\rm ion}^{\rm esc} \dot{N}_i$ respectively, yielding an approximate IGM-to-galaxy mean intensity ratio of ${\overline{I}_{i,\rm IGM}}/{\overline{I}_{i,\rm gal}}\approx{f_{\rm ion}^{\rm esc}}/\left[{f_{\rm ion}^{\rm dust}(1-f_{\rm ion}^{\rm esc})f_i^{\rm dust}}\right]$.

We note that the estimates presented here include only radiative recombination emission. At temperatures above $T\gtrsim10^{4}\,{\rm K}$, collisional excitation can in principle become comparable to or even dominate over radiative recombination. In addition, our estimates rely on recombination emissivities based on Case-B coefficients. In the diffuse IGM, this may not be strictly satisfied (e.g. \citealt{Raga2015}; see also \citealt{Silva2018}). Nevertheless, such deviations are expected to modify the emissivity only at the level of factors of order unity and therefore do not affect our main conclusion. A more realistic treatment, including radiative transfer effects and the inhomogeneous structure of the IGM, will be required for quantitative predictions \citep[e.g.][]{Ambrose2025}.

\subsubsection{LIM Observables with Attenuated and IGM Model}
\label{sec:LIM Observables with Attenuated and IGM Modell}
\begin{figure}
 \includegraphics[width=\columnwidth]{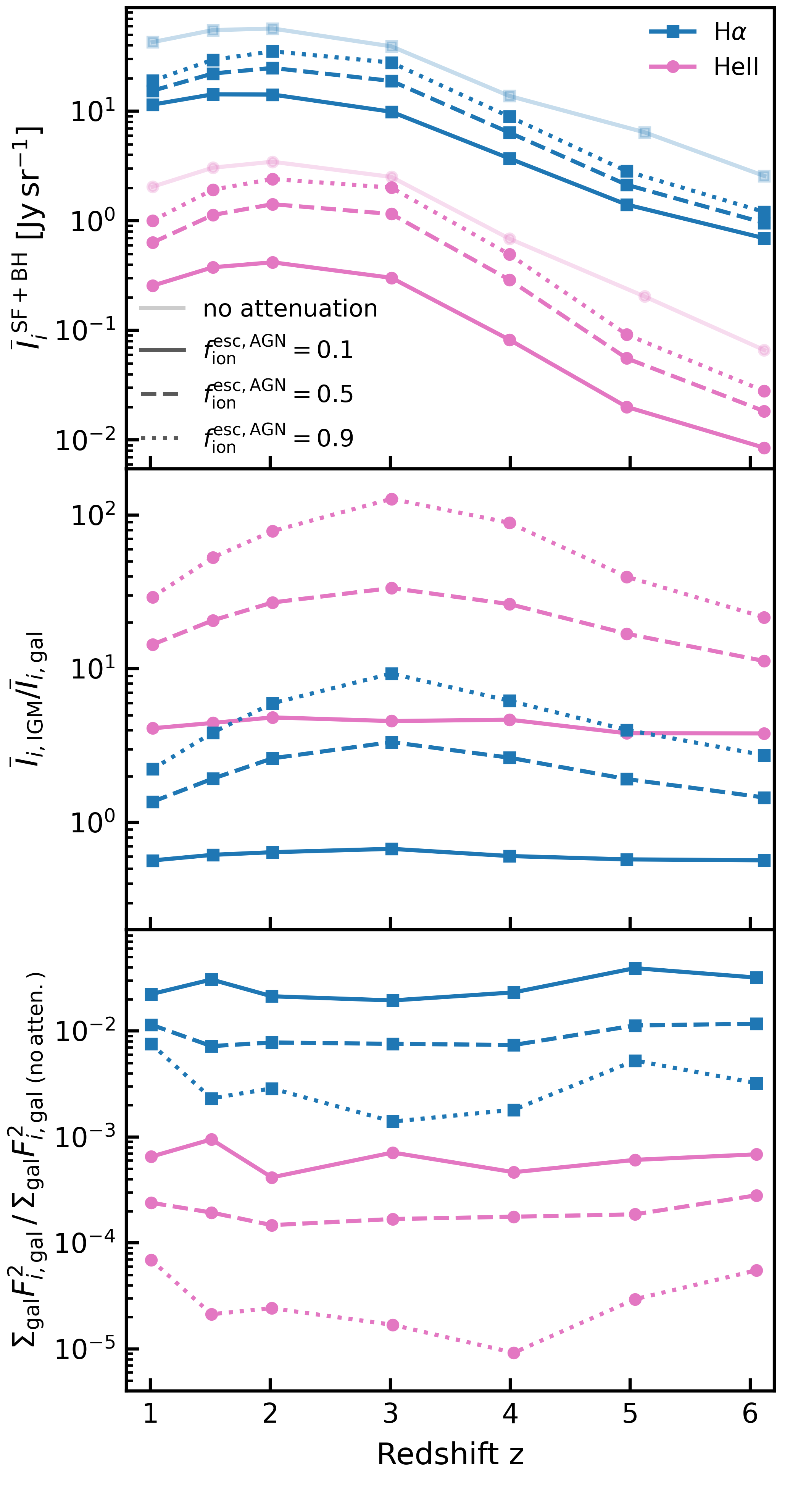}
\caption{
Redshift evolution of the mean intensity and flux-squared statistics for H$\alpha$ (blue) and \ion{He}{II} (magenta), including the effects of dust attenuation, ionizing-photon escape, AGN torus obscuration, and IGM recombination emission. Top panel: total mean intensity, including both galactic and IGM contributions. Middle panel: ratio between the IGM and galactic mean intensities. Bottom panel: ratio of the summed squared fluxes, $\Sigma_{\rm gal} F_{i,\rm gal}^2$, relative to the no-attenuation model, used here as a proxy for the shot-noise contribution. Different line styles correspond to different escape fractions of ionizing photons from AGN, $f_{\rm ion}^{\rm esc,AGN}=0.1$, $0.5$, and $0.9$, while the pale solid curves show the reference no-attenuation model.
}
\label{fig:redshift_torus}
\end{figure}
The top panel of Figure~\ref{fig:redshift_torus} shows the redshift evolution of the mean intensity after including dust attenuation, ionizing-photon escape, AGN torus obscuration, and IGM recombination emission. Compared to the fiducial no-attenuation model, the mean intensity is systematically reduced for both H$\alpha$ and \ion{He}{II}, and the resulting amplitude depends sensitively on the adopted AGN escape fraction $f_{\rm ion}^{\rm esc,AGN}$. As $f_{\rm ion}^{\rm esc,AGN}$ increases, a larger fraction of ionizing photons escapes from galaxies and contributes to recombination emission in the IGM. Consequently, although the galactic component becomes weaker, the total mean intensity partially recovers due to the additional IGM contribution. This effect is more pronounced for \ion{He}{II}, for which the IGM component can exceed the galactic contribution by more than two orders of magnitude at high escape fractions, as shown in the middle panel of Figure~\ref{fig:redshift_torus}. 
We note that the large-scale power scales with the square of the bias-weighted mean intensity of each emitting component. As the IGM generally has a lower bias than galaxies, its relative contribution to the large-scale power is smaller than implied by its contribution to the mean intensity in the middle panel of Figure~\ref{fig:redshift_torus}.

The bottom panel shows the corresponding evolution of the summed squared fluxes, $\Sigma_{\rm gal} F_{i,\rm gal}^2$, normalized by the no-attenuation model, which serves as a simple proxy for the shot-noise contribution. Unlike the mean intensity, the shot-noise proxy is strongly suppressed by attenuation and ionizing-photon escape, because it is dominated by bright compact sources within galaxies rather than diffuse IGM emission. As $f_{\rm ion}^{\rm esc,AGN}$ increases, a larger fraction of ionizing photons escapes into the IGM, reducing the line luminosities of individual galaxies and therefore lowering the shot-noise contribution. 
The suppression is more pronounced for \ion{He}{II} than for H$\rm \alpha$, because \ion{He}{II} emission is more susceptible to dust absorption; the normalized flux-squared sum decreases as large as $\sim 5$ order of magnitude when the escape fraction is high.
By contrast, the H$\alpha$ shot-noise proxy decreases more moderately, reflecting the larger contribution from star-forming galaxies whose escape fractions are fixed to relatively small values in our model. 

In the fiducial no-attenuation model, we demonstrated that SPHEREx and CDIM can detect the H$\alpha$ and \ion{He}{II} power spectra over a wide redshift range, owing in part to the strong shot-noise contribution from BH-powered emission (see Figure~\ref{fig:SN}). The estimates presented in Figure~\ref{fig:redshift_torus} suggest that attenuation and ionizing-photon escape can reduce this signal, thereby modifying the expected detectability.
For H$\alpha$, detection with SPHEREx may become challenging for $f_{\rm ion}^{\rm esc,AGN}\gtrsim0.5$, where the shot noise is suppressed by more than two orders of magnitude, whereas CDIM may remain sensitive even after such a suppression. For \ion{He}{II}, the impact is stronger; for $f_{\rm ion}^{\rm esc,AGN} > 0.1$, the shot noise is reduced by about three orders of magnitude or more, making detection challenging even with CDIM, especially at $z > 4$.
In our simple implementation, the shot noise in the H$\alpha$--\ion{He}{II} cross-power spectrum is given by the square root of the product of the shot-noise terms in the H$\alpha$ and \ion{He}{II} auto-power spectra. For instance, for $f_{\rm ion}^{\rm esc,AGN} = 0.1$, we expect a reduction of $\sim 10^{-2.5}$ in the cross-power shot noise. This suggests that the small-scale cross-power spectrum may remain detectable with CDIM even at $z = 4$--5 and may provide useful constraints on \ion{He}{II} emission.
Nevertheless, these detectability estimates should be interpreted as conservative, since we applied dust attenuation prescriptions calibrated for dusty low-redshift systems.

If the escape fraction of ionizing photons from AGN is substantially higher than that from star-forming regions, the BH-powered contribution would be reduced, which may be advantageous for using LIM to probe the star formation history.
In the H$\alpha$ auto-power spectrum at $z\sim1$, however, even when the AGN escape fraction is increased to $f_{\rm ion}^{\rm esc,AGN}= 0.5$ ($0.9$), the fractional BH contribution to the small-scale power spectrum decreases only slightly, to $P^{\rm BH}/P^{\rm BH+SF}\sim0.7$ ($\sim0.45$) .
We thus conclude that, unless the escape fraction for AGN is quite high (i.e. $f_{\rm ion}^{\rm esc,AGN} > 0.9$), the BH-powered emission dominates in the small-scale power, and therefore the AGN contribution must be removed to inferthe star-formation history as discussed in Section~\ref{sec:AGN_masking}.

\section{Conclusion}
\label{sec:Conclusion}
In this study, we have investigated the contribution of BHs to line-intensity mapping (LIM) signals. 
Focusing on the H$\alpha$ and \ion{He}{II} $1640$\,\AA\ recombination lines, we constructed mock three-dimensional intensity maps using IllustrisTNG simulation. We have quantified the impact of BH-powered emission on a broad set of LIM observables from the mean intensity, voxel intensity distribution (VID), to the three-dimensional auto- and cross-power spectra. Our main results are summarised as follows.

First, BH-powered emission contributes significantly to the global line-intensity budget. For H$\alpha$, while the mean intensity remains dominated by SF-powered emission, the BH contribution can reach $\sim 40$ -- $60$ per cent 
around cosmic noon. Moreover, the \ion{He}{II} mean intensity is even more contributed by BH-powered emission over a wide redshift range, with typical BH fractions of $\sim 60$ -- $80$ per cent. Despite this large luminosity-weighted contribution, the spatial filling factor of BH-dominated regions is small. Only a few per cent of voxels are dominated by BH-powered emission for either line at all redshifts considered.
We have further shown that BH-powered emission leaves a distinct signature on the one-point statistics of the intensity field, especially for \ion{He}{II}. 
We note that our adopted stellar \ion{He}{II} production efficiency likely represents an approximate upper limit, as we adopt values near the upper end of those inferred from observations and population synthesis models.
Our estimated contribution of BH-powered emission relative to the SF-powered component in \ion{He}{II} may therefore be underestimated.
 
Second, we find that BH-powered emission strongly enhances the power spectra, especially in the shot-noise–dominated regime at small scales. This behaviour arises because the shot-noise term is weighted by the square of the source luminosity, and is therefore highly sensitive to rare but extremely luminous BH-powered emitters. Even when the BH contribution to the mean intensity is modest, the fractional contribution to the shot-noise term can approach unity. This effect is particularly pronounced for \ion{He}{II}, for which the small-scale power spectrum is largely governed by BH-powered emission.

Third, we have assessed the observational implications for forthcoming LIM surveys. For a SPHEREx-like configuration, the bright end of the H$\alpha$ VID at $z \lesssim 4$ is accessible and corresponds to a regime in which BH-powered emission contributes more than half of the total voxel intensity. In contrast, the intrinsic surface brightness of \ion{He}{II} is too low for SPHEREx to probe the BH-dominated bright tail. For a CDIM-like experiment with higher angular and spectral resolution and improved sensitivity, both the intermediate-intensity H$\alpha$ regime and the BH-dominated \ion{He}{II} bright tail become accessible. Our results therefore indicate that future high-sensitivity LIM surveys offer a unique opportunity to directly probe BH-driven line emission in a statistical manner, beyond what is possible with current-generation instruments.
We note that probing \ion{He}{II} emission at $z\lesssim4$ requires wavelength coverage extending into the UV–optical regime, beyond the wavelength ranges of current LIM survey concepts. We therefore introduced a hypothetical auxiliary short-wavelength extension in our analysis to explore the detectability of low-redshift \ion{He}{II} emission.

We have also examined the impact of dust extinction and the escape of ionizing photons. These attenuation effects reduce the overall signal levels, but our main qualitative conclusions regarding the relative roles of BH- and SF-powered emission remain unchanged.

We note that our analysis involves several simplifying assumptions.
For instance, the ionizing spectra of both stellar populations and accreting BHs are modelled using simple prescriptions.
We also assume spatially uniform and redshift-independent dust attenuation. These assumptions should be sufficient to examine the critical role of BH-powered emission in LIM statistics, but more realistic modelling including galaxy-type dependent attenuation, AGN obscuration, and foreground-mitigation pipelines will be required for robust survey predictions.

Overall, our results suggest that accreting BHs constitute a previously underexplored but potentially dominant component of LIM signals, particularly for emission lines sensitive to hard ionizing radiation such as \ion{He}{II}. 
Similar BH-powered enhancements is also expected for other recombination lines whose luminosities trace the ionizing photon budget, although the effect may be less pronounced or more model-dependent for metal lines (see Appendix~\ref{sec:Line Contamination}). 
Even a sparse population of luminous BHs can strongly modify both the one-point and two-point statistics of the intensity field. Explicitly incorporating BH-powered emission is therefore essential for a physically complete interpretation of LIM measurements, especially in studies aiming to connect LIM observables to the cosmic history of star formation and BH growth. 
Future studies incorporating more detailed and accurate physical models of galaxies and BHs and larger simulation volumes will improve our understanding of LIM signals. 
Fully exploiting the scientific potential of future LIM missions will require statistical studies based on cross-correlation functions.


\section*{Acknowledgements}

The authors acknowledge financial support from Japan Society for the Promotion of Science (JSPS) KAKENHI Grant Numbers JP23K03446, JP23K20035, JP24H00004, JP26K17192.

\section*{Data Availability}

The TNG simulation data used in this study is publicly available.\footnote{https://www.tng-project.org}
The mock data used in this analysis and the corresponding outputs are available upon reasonable request.



\bibliographystyle{mnras}
\bibliography{example} 




\appendix

\section{Comparison with Observations}
\label{sec:Sanity check}

\subsection{H$\alpha$ Luminosity Function}
\label{sec:lf_Halpha}
\begin{figure*}
\centering

\includegraphics[width=0.95\textwidth]{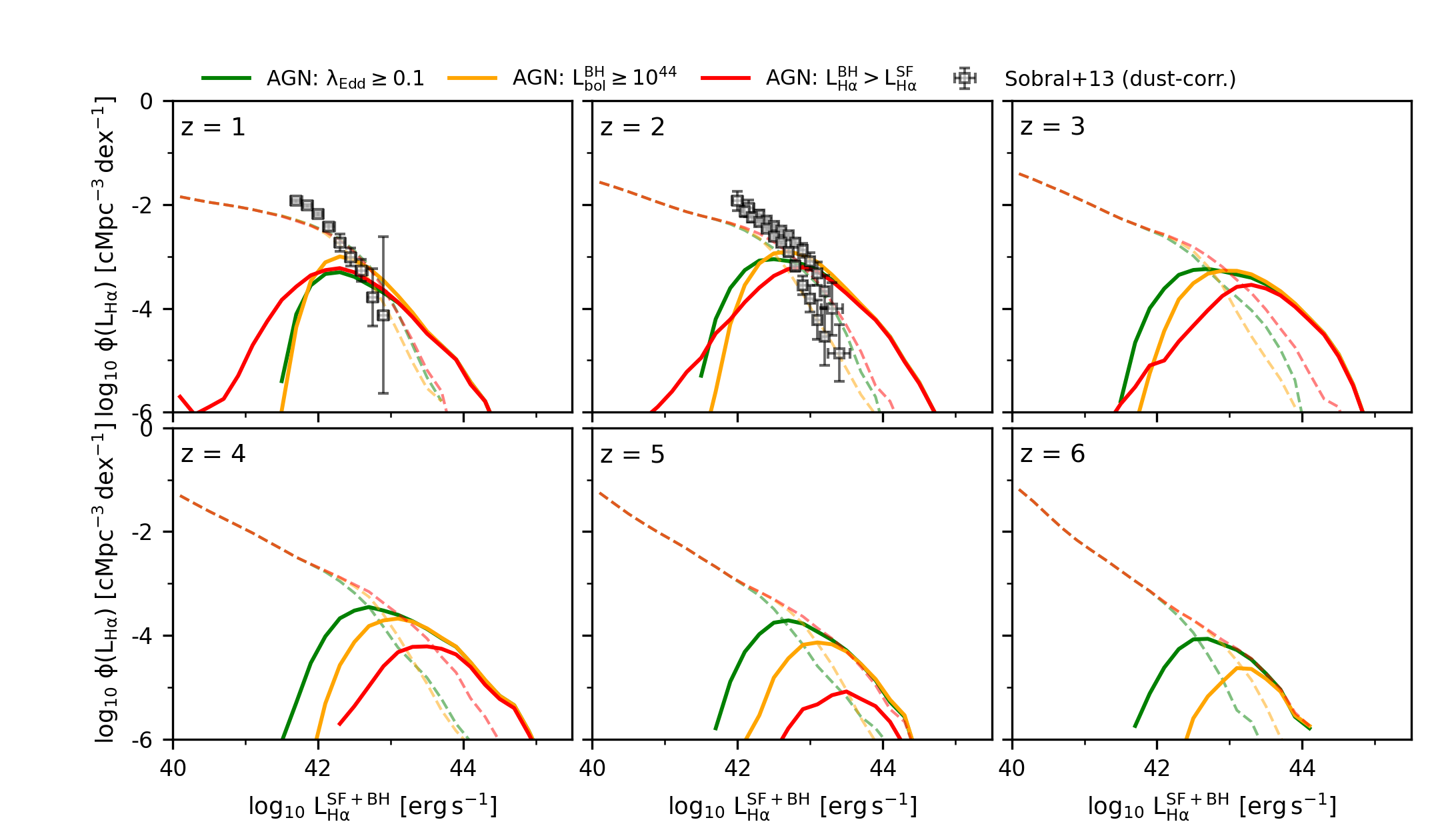}

\vspace{3mm}

\includegraphics[width=0.95\textwidth]{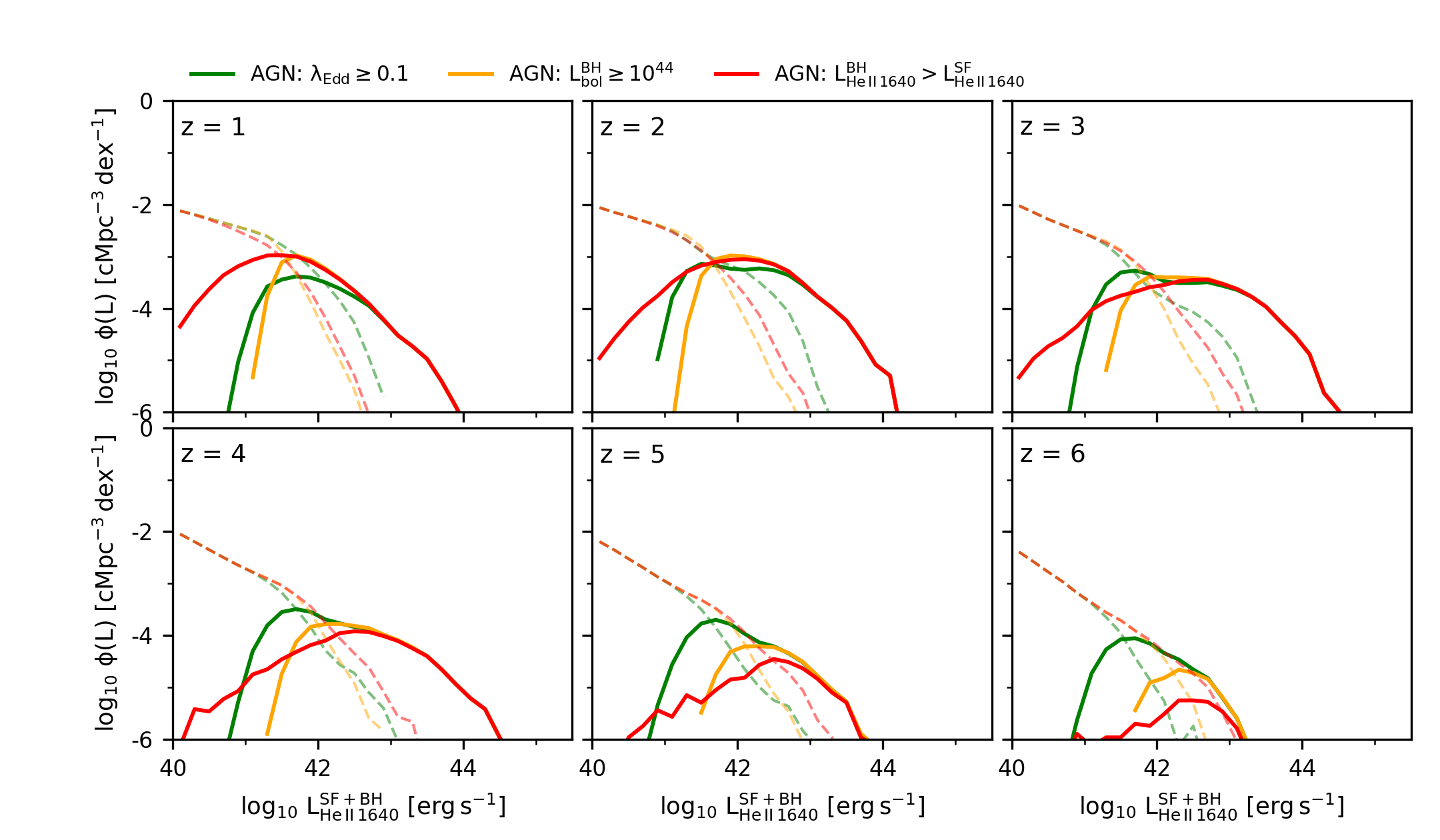}

\caption{
Luminosity functions derived from the TNG300 mock galaxy catalog under three different AGN definitions.
Top panels show the H$\alpha$ luminosity functions, while bottom panels show the \ion{He}{II} $1640$ \AA\ luminosity functions. Green, orange, and red curves correspond to AGN-selected subsamples defined by (i) an Eddington ratio threshold, $\lambda_{\rm Edd}\geq0.1$, (ii) a bolometric luminosity cut, $L_{\rm bol}^{\rm BH}\geq10^{44}\,{\rm erg\,s^{-1}}$, and (iii) line-luminosity dominance, $L_{\rm line}^{\rm BH}>L_{\rm line}^{\rm SF}$, respectively. For each definition, solid curves show the AGN-selected subsample, while dashed curves of the same colour show the complementary non-AGN galaxy population.
Observational measurements from \citet{Sobral2013} are shown in the $z=1$ and $z=2$ H$\alpha$ panels (dust-corrected values), where the $z=2$ panel includes data at $z\simeq1.47$ and $z\simeq2.23$.
}
\label{fig:LF_multiAGN}
\end{figure*}

We validate our line-luminosity modelling by calculating the H$\alpha$ luminosity functions (LFs). 
The top panel of Figure~\ref{fig:LF_multiAGN} shows our H$\alpha$ LFs at $z=1-6$. The H$\alpha$ luminosity of each galaxy is computed as the sum of the SF– and BH–powered components. 
Since observational LFs are generally constructed separately for AGN and non-AGN populations, we divide our simulated galaxy sample in the same way. We consider three different AGN identification criteria:
(i) an Eddington ratio threshold, $\lambda_{\rm Edd}\geq0.1$, (ii) a bolometric luminosity cut, $L_{\rm bol}^{\rm BH}\geq10^{44}\,{\rm erg\,s^{-1}}$, and (iii) a criterion based on H$\alpha$ dominance, $L_{\rm H\alpha}^{\rm BH}>L_{\rm H\alpha}^{\rm SF}$.
The solid and dashed lines represent the LFs of AGN and non-AGN populations in our sample.
We overplot dust-corrected H$\alpha$ LFs of star-forming galaxies (non-AGN) from \citet{Sobral2013} at $z\simeq0.8$, $1.47$, and $2.23$.

We confirm that the predicted H$\alpha$ LFs are broadly consistent with the available observational constraints. 
On the faint side, where the emission is dominated by star formation, the H$\alpha$ LFs closely follow the dust-corrected measurements of star-forming galaxies from \citet{Sobral2013} at $z\simeq0.8-2.23$, largely independent of the adopted AGN definition. We verify that the remaining discrepancy has only a minor impact on the luminosity-weighted integral (i.e. the mean intensity), resulting in a bias of at most a few per cent.

We also present the predicted \ion{He}{II} LFs in the bottom panel of Figure~\ref{fig:LF_multiAGN}, as a reference for future observational studies, where the emission is expected to be more strongly influenced by BH-powered ionizing radiation.

\subsection{AGN Fraction as a Function of H$\alpha$ Luminosity}
\label{sec:AGN Fraction as a Function of Ha Luminosity}
\begin{figure*}
 \includegraphics[width=\textwidth]{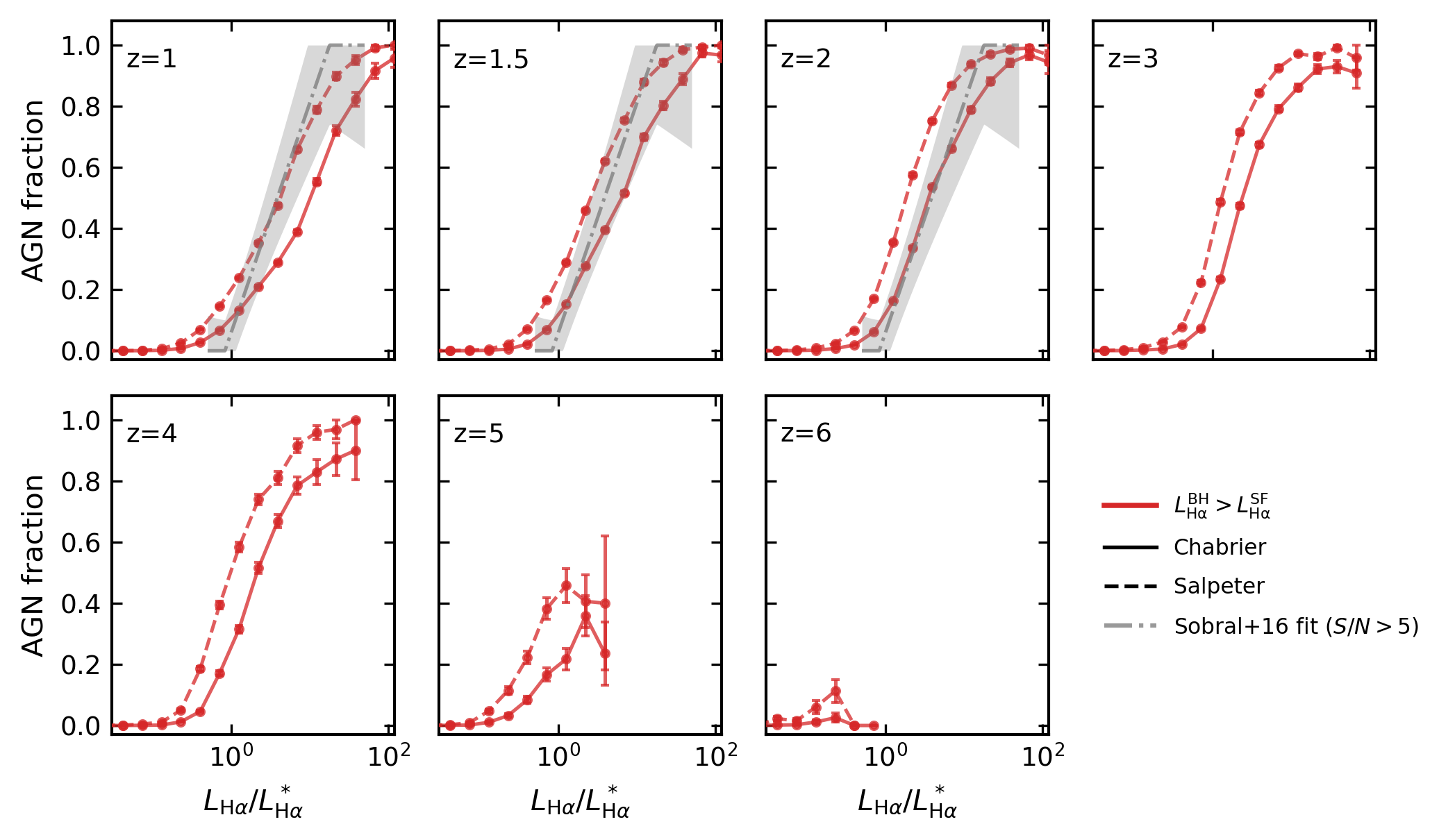}
\caption{
AGN fraction as a function of H$\alpha$ luminosity normalized by the characteristic luminosity, $L_{\mathrm{H}\alpha}/L_{\mathrm{H}\alpha}^{*}$, at $z=1-6$. The AGN number fraction is defined as the fraction of galaxies satisfying $L_{\mathrm{H}\alpha}^{\rm BH} > L_{\mathrm{H}\alpha}^{\rm SF}$ within each luminosity bin. Solid and dashed red curves show results obtained using the Chabrier and Salpeter IMF calibrations, respectively. The grey dashed line shows the best-fitting relation for the $S/N>5$ sample from \citet{Sobral2016}, and the grey shaded region indicates the corresponding $1\sigma$ uncertainty. Error bars on our model results show the binomial uncertainties, $\sigma_f=[f_{\rm AGN}(1-f_{\rm AGN})/N_{\rm gal}]^{1/2}$, where $f_{\rm AGN}$ is the AGN fraction and $N_{\rm gal}$ is the number of galaxies in each luminosity bin.
}
\label{fig:AGN_fraction_Sobral}
\end{figure*}

We compare the predicted AGN number fraction with the observational estimates reported by \citet{Sobral2016}. In \citet{Sobral2016}, AGN are identified through either broad H$\rm \alpha$ emission or line ratios. 
This selection can be approximately interpreted as selecting systems in which the BH-powered H$\rm \alpha$ luminosity exceeds the star-formation-powered contribution, i.e. $L_{\mathrm{H}\alpha}^{\rm BH} > L_{\mathrm{H}\alpha}^{\rm SF}$.
We therefore use this luminosity criterion as our definition of AGN here.

Figure~\ref{fig:AGN_fraction_Sobral} shows the AGN number fraction as a function of H$\alpha$ luminosity normalized by the characteristic luminosity, $L_{\mathrm{H}\alpha}/L_{\mathrm{H}\alpha}^{*}$, over $z=1-6$. 
The characteristic luminosity is computed using the empirical relation from \citet{Sobral2013}, $\log L_{\mathrm{H}\alpha}^{*} = 0.45 z + 41.47$.
Our results reproduce the qualitative trend reported by \citet{Sobral2016}: the AGN fraction increases rapidly toward the bright end of the H$\alpha$ luminosity function. The predicted AGN fractions rise from $\sim10-20\%$ around $L_{\mathrm{H}\alpha}^{*}$ to $\sim30-60\%$ at $\sim5L_{\mathrm{H}\alpha}^{*}$, and reach $\gtrsim90\%$ at $\sim50L_{\mathrm{H}\alpha}^{*}$. This is broadly consistent with the observational estimates of \citet{Sobral2016}, who reported AGN fractions of $\sim10$--$15\%$, $\sim25\%$, $\sim50\%$, and nearly $100\%$ at $L_{\mathrm{H}\alpha}^{*}$, $\sim2L_{\mathrm{H}\alpha}^{*}$, $\sim5L_{\mathrm{H}\alpha}^{*}$, and $\sim50L_{\mathrm{H}\alpha}^{*}$, respectively. The comparison is only possible at $z\lesssim2$, but our model predicts a similar monotonic increase toward higher luminosities at higher redshifts.

\subsection{Bolometric Luminosity Function of Central BHs}
\label{sec:lf_Lbol}
\begin{figure*}
 \includegraphics[width=\textwidth]{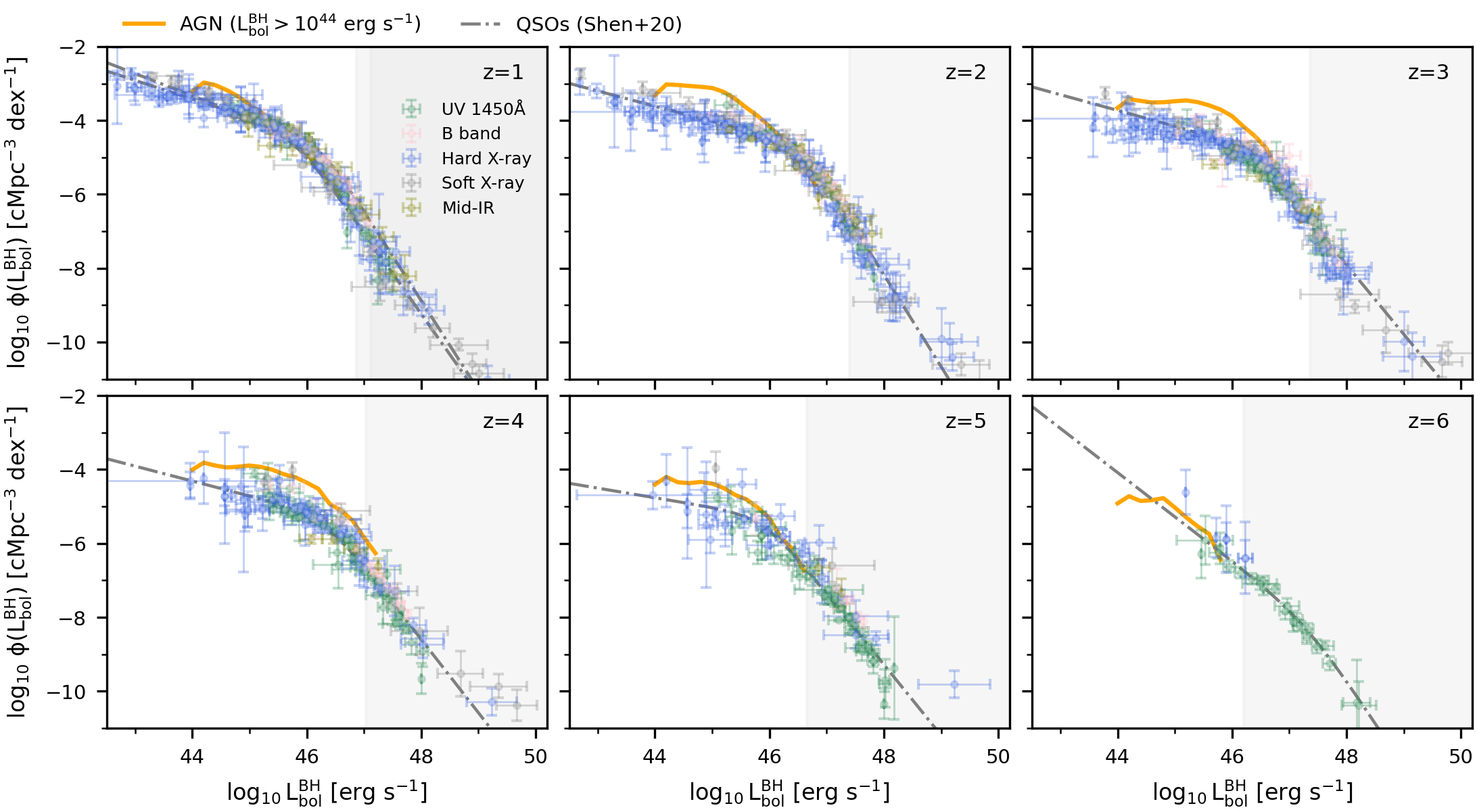}
 \caption{
Bolometric luminosity function (LFs) of central BHs at $z=1$–$6$, derived from our mock galaxy catalogue. Only galaxies hosting central BHs with bolometric luminosities $L_{\rm bol}^{\rm BH}>10^{44}{\rm erg ~s^{-1}}$ are included, thus restricting the sample to the luminous AGN-like regime. Observed quasar luminosity function (QLFs) data points in the UV, optical (B band), hard/soft X-ray, and mid-infrared bands, converted to bolometric luminosities following \citet{Shen2020}, are shown for comparison. The grey dash-dotted curves also show the bolometric QLFs obtained by fitting these multi-wavelength measurements with the parametric 'free' model of \citet{Shen2020}; for $z=1$ we plot the two closest redshift bins ($z=0.8$ and $z=1.2$), as no QLF is tabulated at exactly this redshift. The grey shaded regions indicate luminosities for which the expected number of objects in TNG300 simulation volume is smaller than unity, $N(L_{\rm bol}^{\rm BH})=\phi(L_{\rm bol}^{\rm BH})\,V_{\rm sim}\,\Delta\log_{10}L_{\rm bol}^{\rm BH}<1$, highlighting the range in which the bright end of the observed QLF cannot be sampled due to the finite simulation volume. Within the statistically sampled regime ($N\gtrsim 1$), our results show good agreement with the observed QLFs, confirming the validity of our bolometric luminosity computation.
 }
 \label{fig:LF_Lbol}
\end{figure*}
To validate our implementation of bolometric luminosity calculations for central BHs, we examine the bolometric luminosity function (LFs) derived from our mock galaxy catalogue. Here, we choose only systems with $L_{\rm bol}^{\rm BH}>10^{44}{\rm erg~s^{-1}}$ in order to focus on the luminous AGN-like population that can be meaningfully compared with observational quasar luminosity functions (QLFs).

Figure~\ref{fig:LF_Lbol} shows the resulting bolometric LFs at $z=1$-$6$ with solid lines, together with multi-wavelength QLF measurements (X-ray, UV, optical, and mid-infrared) converted to bolometric luminosities following the prescription of \citet{Shen2020}. 
The grey dash-dotted curves show the best-fit bolometric QLFs obtained by applying the parametric free model of \citet{Shen2020}. 
We also indicate, with grey shading, luminosity range for which the expected number of AGN within the TNG300 volume falls below unity. 

Comparing the simulated and observed LFs, we find that at low redshift ($z\sim1$) the simulation agrees well with the observed QLFs. 
At higher redshift ($z\gtrsim2$), however, the simulation predicts a somewhat larger AGN abundance than inferred from the observed QLFs. The same trend has been pointed out in the original IllustrsTNG paper \citep{Weinberger2017}. 

The excess in our model is apparent at the faint end of the QLFs. 
An important caveat, however, is that the abundance of Compton-thick AGN remains highly uncertain, especially at low luminosities \citep{Buchner2015}. 
Such heavily obscured populations are likely to be underrepresented, or even missed, in observed QLFs. 
Recent observations have also revealed that a number of heavily obscured AGN candidates, including so-called little red dots (LRDs), exhibit strong H$\alpha$ emission despite their obscured nature \citep[e.g.][]{Matthee2024}. 
Therefore, the discrepancy at the faint end does not necessarily imply that our model overestimates the true AGN abundance or their H$\alpha$ output. Rather, it may reflect the incompleteness of current observational constraints on obscured AGN populations and the uncertainties in converting observed quantities to intrinsic bolometric and H$\alpha$ luminosities. 
We therefore retain our fiducial AGN model throughout the analysis.


\section{Line interlopers}
\label{sec:Line Contamination}
\begin{figure}
 \includegraphics[width=\columnwidth]{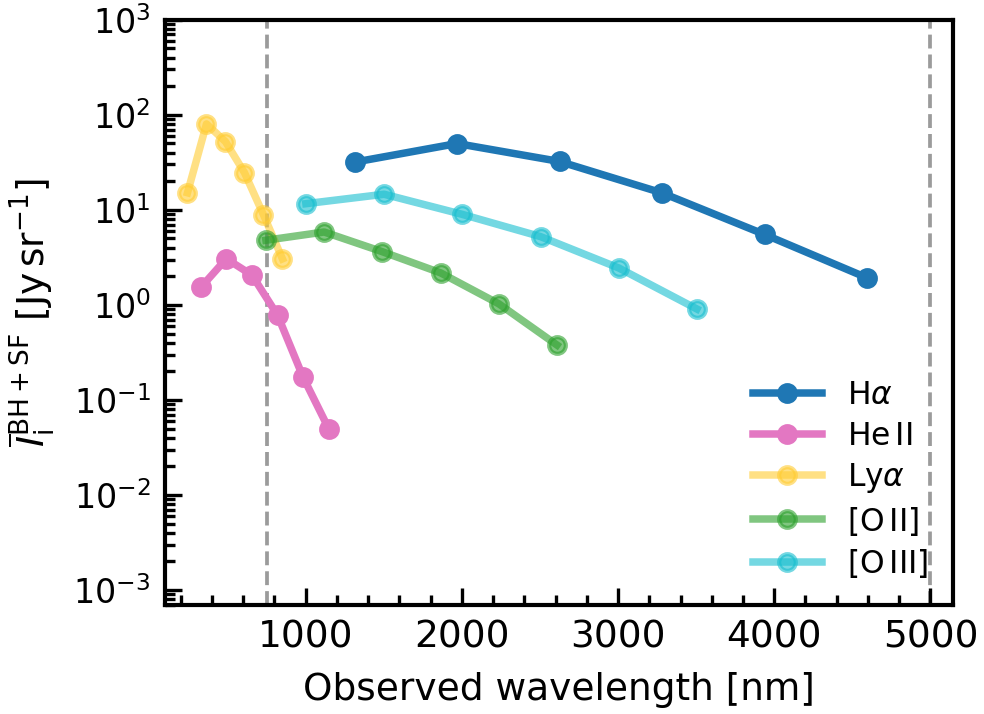}
\caption{
Mean intensities of five emission lines as a function of observed wavelength. The upper and lower panels show the results without and with attenuation, respectively. Coloured curves correspond to H$\alpha$ 6562 (blue), \ion{He}{II} 1640 (magenta), Ly$\alpha$ 1216 (yellow), [\ion{O}{II}] 3727 (green), and [\ion{O}{III}] 5007 (cyan). The vertical dashed lines mark the wavelength coverage of the SPHEREx mission (750–5000 nm). The mean intensity is averaged over all SPHEREx-like voxels at each redshift bin. For each curve, the data points are ordered from left to right in increasing redshift, corresponding to $z=1$ to $z=6$.}
\label{fig:obswave_meanint}
\end{figure}

As the observed signal includes emission from all unresolved sources along the line of sight, line contamination is a fundamental challenge in LIM surveys. If the contribution from interlopers is comparable to or exceeds that of the target line, contaminated voxels must be removed from the observational maps. In the case of H$\alpha$ and \ion{He}{II} intensity mapping, several bright interlopers are expected to contribute significantly, making contamination removal particularly important. In this section, we model the intensities of the emission lines that act as interlopers for H$\alpha$ and \ion{He}{II}. 

For H$\alpha$ and \ion{He}{II} intensity mapping, the observed signal is subject to contamination from several strong optical and UV emission lines, such as the ionized oxygen [\ion{O}{II}] 372.7 nm and [\ion{O}{III}] 500.7 nm lines, and the hydrogen H$\beta$ 486.1 nm and Ly$\alpha$ 121.6 nm lines. Additional contamination may arise from the [\ion{N}{II}] 6548/6583 nm and [\ion{S}{II}] 6717/6731 nm doublets. Among these interlopers, [\ion{O}{III}], [\ion{O}{II}], and Ly$\alpha$ are expected to provide the largest contributions owing to their high intrinsic luminosities and abundance. Here, we therefore focus on these dominant contaminants.
For [\ion{O}{III}] and [\ion{O}{II}], we adopt empirical luminosity relations calibrated from observations, because the luminosities of these forbidden metal lines cannot be robustly determined from the ionizing photon budget alone and depend strongly on the physical conditions of the emitting gas.
For simplicity, we restrict our analysis to the Chabrier IMF.

\paragraph*{[\ion{O}{III}$]\,\boldsymbol{500.7}$ nm}
The [\ion{O}{III}] doublet at 495.9 nm and 500.7 nm has an intrinsic intensity ratio of [500.7 nm]/[495.9 nm] $\sim$3. 
We treat the combined emission from these two lines as our [\ion{O}{III}] luminosity estimator.
For the SF-powered \ion{O}{III} emission, we adopt the empirical relation from \citet{Ly2007}: 
\begin{align}
L_{[\mathrm{O\,III}]}^{\mathrm{SF}} = 1.3\times10^{41}
\left(\frac{\mathrm{SFR}}{M_\odot\,\mathrm{yr}^{-1}}\right)\,\mathrm{erg\,s^{-1}}.
\end{align}
For the BH component, we infer the [\ion{O}{III}] luminosity from the observed correlation between hard X-ray (2–10 keV) luminosity, $L_X^{\rm BH}$, and [\ion{O}{III}] line luminosity \citep{Heckman2005, Panessa2006},
\begin{align}
\log_{10}L_{[\mathrm{O\,III}]}^{\mathrm{BH}} = 
\frac{7.3 + \log_{10}L_X^{\mathrm{BH}}}{1.2}~\mathrm{erg\,s^{-1}}.
\end{align}
We adopt the hard X-ray bolometric correction of quasars from \citet{Shen2020}, 
\begin{equation}
\frac{L_{\rm bol}^{\rm BH}}{L_{X}^{\rm BH}}
=
c_1
\left(
\frac{L_{\rm bol}^{\rm BH}}{10^{10}L_\odot}
\right)^{k_1}
+
c_2
\left(
\frac{L_{\rm bol}^{\rm BH}}{10^{10}L_\odot}
\right)^{k_2},
\end{equation}
where the best-fitting parameters are $(c_1, k_1, c_2, k_2) = (4.073, -0.026, 12.60, 0.278)$ from their Table~1. 

\paragraph*{[\ion{O}{II}]$\,\boldsymbol{372.7}$ nm}
For the SF-powered [\ion{O}{II}]~$372.7$ nm line emission, the luminosity is estimated by 
combining the observed [\ion{O}{II}]/H$\alpha$ ratios with the H$\alpha$–SFR conversion derived from population synthesis models \citep[e.g.][]{Kennicutt1998}.
This gives the following relation:
\begin{align}
L_{[\mathrm{O\,II}]}^{\mathrm{SF}} = 7.1\times10^{40}\left(\frac{\mathrm{SFR}}{M_\odot\,\mathrm{yr}^{-1}}\right)\,\mathrm{erg\,s^{-1}}.
\end{align}
Observational studies show that the [\ion{O}{II}] strength of AGN typically amounts to $\sim$10--30 per cent of that of [\ion{O}{III}] line \citep{Ferland1986, Ho2005, Kim2006}. We model the BH-powered [\ion{O}{II}] emission by adapting the empirical best-fit relation reported by \citet{Kim2006}, which is based on a sample of approximately 3600 type 1 AGN at $z<0.3$ selected from the Sloan Digital Sky Survey (SDSS):
\begin{align}
\log_{10}L_{[\mathrm{O\,II}]}^{\mathrm{BH}} 
= 26 + 0.36\,\log_{10}L_{\mathrm{O\,III}}^{\mathrm{BH}}.
\end{align}
We note that \citet{Kim2006} conclude that the [\ion{O}{II}] emission in their sample is predominantly produced by AGN photoionization, with little contribution from star formation activity. Their results therefore provide an appropriate calibration dataset for our analysis of BH-driven [\ion{O}{II}] emission \citep[see also][for more details]{Silverman2009}.

\paragraph*{Ly$\boldsymbol{\alpha}$}
The hydrogen Ly$\alpha$ line is the most energetic hydrogen emission line from galaxies, with a rest-frame ultraviolet wavelength of 121.6 nm. We estimate the intrinsic Ly$\alpha$ luminosity of individual sources by adopting a Ly$\alpha$/H$\alpha$ line ratio of 8.7, as expected from case-B recombination theory \citep{Brocklehurst1971}. 

\vspace{1.0em}
In Figure~\ref{fig:obswave_meanint}, we show the intrinsic mean intensities of the target lines H$\alpha$ and \ion{He}{ii}, together with the interloping lines Ly$\alpha$, [\ion{O}{ii}], and [\ion{O}{iii}], as a function of the observed wavelength. 
The mean intensity of H$\alpha$ generally exceeds those of the main optical interlopers, [\ion{O}{ii}] and [\ion{O}{iii}]. By contrast, the \ion{He}{ii} line is intrinsically much fainter, and its predicted mean intensity is often comparable to, or smaller than, those of several interloping lines. 

The relative contribution of accreting BHs to the total mean intensity also varies among different emission lines. 
By construction, the BH contribution to Ly$\alpha$ is the same as that to H$\alpha$, reaching $30-50$ per cent at $z=1-3$. In contrast, for the metal lines considered here, the BH contribution remains significantly smaller: $\sim5-8$ per cent for [\ion{O}{iii}] and $\leq 5$ per cent for [\ion{O}{ii}] over the same redshift range, decreasing toward higher redshift. 
The smaller BH contribution to the forbidden metal lines is qualitatively consistent with the fact that [\ion{O}{iii}] and [\ion{O}{ii}] are predominantly narrow-line-region tracers, whereas H$\alpha$ can include both narrow and broad components in unobscured AGN. We do not explicitly decompose the AGN emission into broad- and narrow-line components, but the use of empirical relations based on total line luminosities can statistically encode this difference.

These results indicate that the line interlopers can be non-negligible in realistic observations and should be carefully treated when extracting the target LIM signal.
A variety of mitigation strategies have been developed in LIM studies, including masking of bright foreground galaxies using ancillary survey data \citep[e.g.][]{Sun2018, VanCuyck2023}, anisotropy in the power-spectrum \citep[e.g.][]{Lidz2016,Cheng2016}, cross-correlation \citep[e.g.][]{Roy2024,Bernal2025}, sparse modelling \citep{Cheng2020}, and machine-learning techniques \citep[e.g.][]{Moriwaki2021}. 
Incorporating such techniques into the modelling and inference framework is an important direction for future work.

\section{Voxel Intensity Distribution for CDIM}
\label{sec:VIDs for CDIM}
\begin{figure*}
 \includegraphics[width=\textwidth]{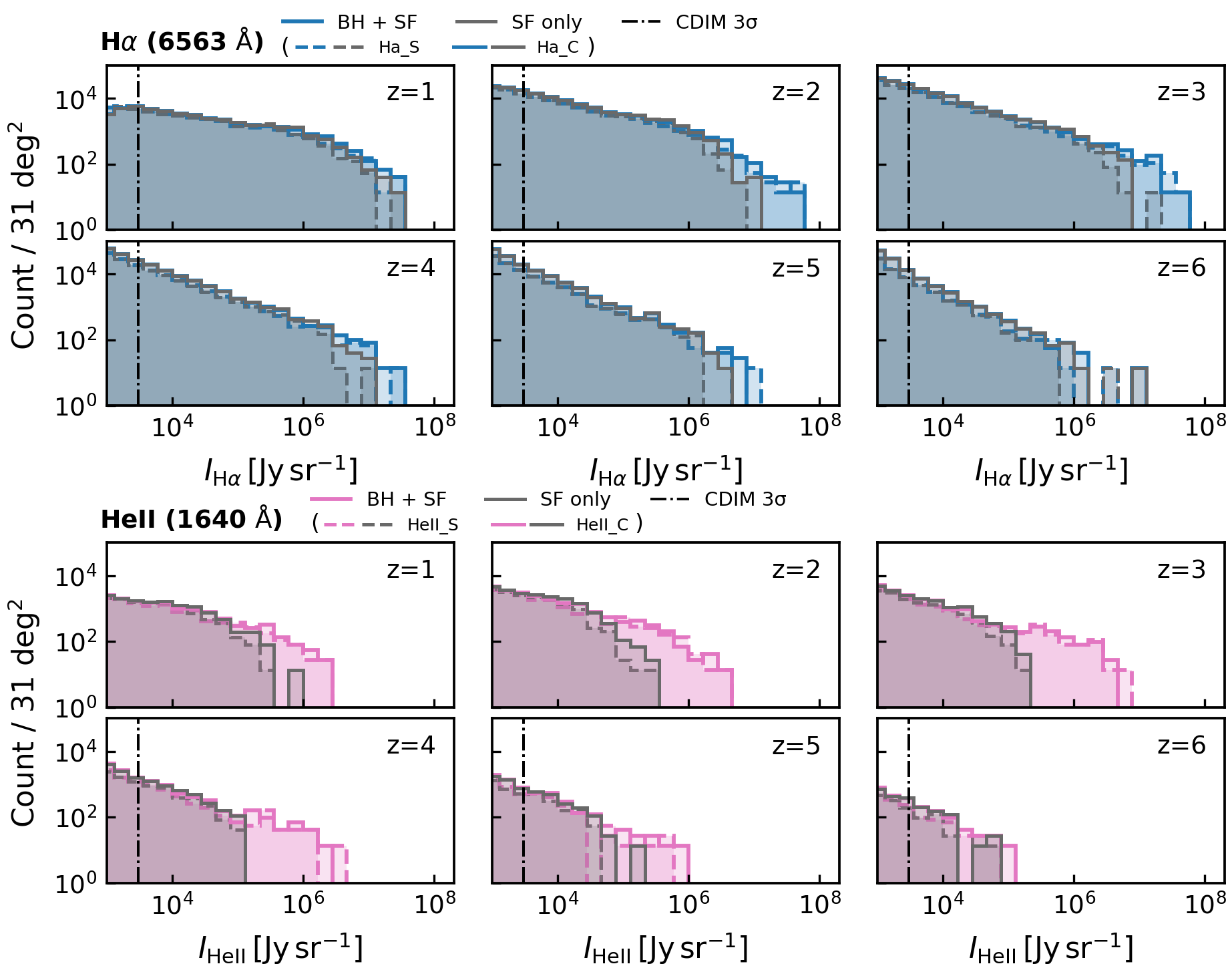}
 \caption{
 Same as Figure~\ref{fig:VID_count_Ha_HeII}, but for CDIM-like voxelization with an angular resolution of $1''$ and spectral resolution $R=500$. The distributions are scaled to an effective survey area of 31 deg$^{2}$, corresponding to the fiducial CDIM survey area. Vertical dash-dotted lines indicate the adopted CDIM $3\sigma$ surface-brightness thresholds, assuming $I_\nu = 10^{-20}\,{\rm erg\,s^{-1}\,cm^{-2}\,sr^{-1}\,Hz^{-1}}$.
}
 \label{fig:VID_count_Ha_HeII_CDIM}
\end{figure*}
Figure~\ref{fig:VID_count_Ha_HeII_CDIM} shows the voxel intensity distributions for a CDIM-like survey configuration with angular resolution $1''$ and spectral resolution $R=500$. Compared to the SPHEREx-like case shown in Figure~\ref{fig:VID_count_Ha_HeII}, the improved angular and spectral resolution extends the VID toward higher intensities and allows substantially fainter regions of the distribution to be probed relative to the instrumental sensitivity threshold.

For H$\alpha$, CDIM can access not only the BH-dominated bright end of the VID but also intermediate-intensity regimes where the emission is predominantly produced by star-forming galaxies. For \ion{He}{II}, the improved sensitivity allows the high-intensity tail of the VID to be detected over a broad redshift range. In this regime, BH-powered emission produces a clear excess relative to the SF-only prediction, indicating that CDIM-like LIM observations may provide a sensitive probe of BH-driven \ion{He}{II} emission.

\section{Voxel Anisotropy}
\label{sec:Voxel Anisotropy}
\begin{figure}
 \includegraphics[width=\columnwidth]{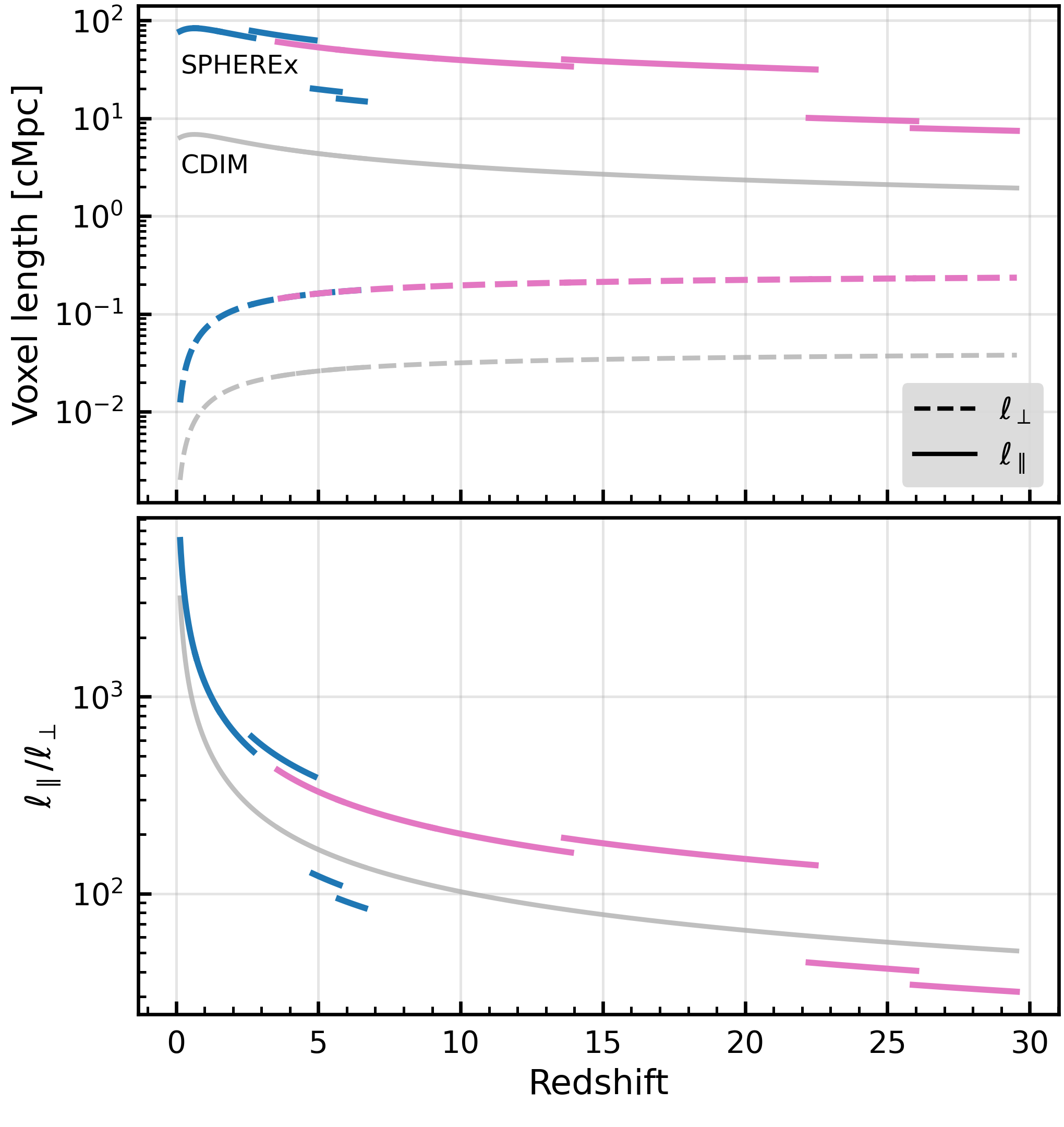}
 \caption{
 Comoving voxel lengths perpendicular and parallel to the line of sight, $\ell_{\perp}$ (dashed) and $\ell_{\parallel}$ (solid), as a function of redshift. The coloured curves show the SPHEREx configuration for the H$\alpha$ (blue) and \ion{He}{II}~1640 (pink) emission lines, adopting the band-dependent spectral resolving power and an angular pixel size of $6.2^{\prime\prime}$. The grey curves indicate the CDIM improved configuration, assuming a uniform spectral resolving power of $R=500$ and an angular pixel size of $1^{\prime\prime}$. The lower panel shows the corresponding anisotropy ratio $\ell_{\parallel}/\ell_{\perp}$.
 }
 \label{fig:voxel_volume_anisotropy}
\end{figure}
LIM observations measure the sky brightness as a function of observing frequency and angular position on the sky, forming a three-dimensional intensity field. In an idealized case, where the intensity field is defined in a uniform comoving volume and observational or projection effects are absent, the three-dimensional power spectrum of line-intensity fields is expected to be statistically isotropic, 
and therefore depends only on the magnitude of the wavevector, $k = |\mathbf{k}|$.

In practical LIM observations, however, various observational effects introduce anisotropy between the transverse and line-of-sight directions. Such anisotropies have been widely discussed in the LIM literature, primarily in the context of line-confusion (interloper) projection effects \citep{Visbal2010,Gong2014,Cheng2016,Lidz2016}. Here, we do not disucuss interloper projection effects. We focus on a more instrumental but practically important source of anisotropy that arises from the voxelization of the three-dimensional intensity cube itself. In particular, the transverse and line-of-sight directions generally have different spatial resolutions and extents, which can introduce artificial anisotropies in Fourier space and affect the sampling of the three-dimensional power spectrum.

To quantify the anisotropy introduced by the voxelization, we define the comoving voxel sizes along and perpendicular to the line of sight, $\ell_{\parallel}$ and $\ell_{\perp}$, respectively. For a voxel with an observed angular size $\Delta\theta$ and an observed frequency width $\Delta\nu_{\rm obs}$ at redshift $z$, the voxel sizes are given by 
\begin{align}
\label{eq:l_parallel_perp}
\ell_{\parallel}(z)
&\simeq
\frac{d\chi}{d\nu_{\rm obs}}\,\Delta\nu_{\rm obs}
=
\frac{c\,(1+z)^2}{H(z)\,\nu_{\rm rest}}\,\Delta\nu_{\rm obs},\\
\ell_{\perp}(z)
&=
D_{\rm A}(z)\,\Delta\theta ,
\end{align}
where $\chi$ is the radial comoving distance, $H(z)$ is the Hubble parameter, $\nu_{\rm rest}$ is the rest-frame frequency of the emission line, $\nu_{\rm obs}$ denotes the observed frequency, and $D_{\rm A}(z)$ is the comoving angular diameter distance. Here we approximate the line-of-sight voxel size using the local derivative, which is applicable when the variation of $d\chi/d\nu_{\rm obs}$ across a single frequency channel of width $\Delta\nu_{\rm obs}$ is small.
The ratio between the line-of-sight and transverse voxel sizes, $\ell_{\parallel}/\ell_{\perp}$, 
provides a measure of the anisotropy of the voxel geometry and determines how the Fourier-space sampling differs between the parallel and transverse directions.

Figure~\ref{fig:voxel_volume_anisotropy} illustrates the resulting voxel geometry for the SPHEREx and CDIM configurations considered in this work. In the main text, we assume a fixed spectral resolution $R$, whereas here we show values obtained using the actual wavelength-dependent SPHEREx resolution. 
In both SPHEREx and CDIM, the line-of-sight voxel length is much larger than the transverse size, leading to a strongly anisotropic voxel geometry of $\ell_{\parallel}/\ell_{\perp} \sim 10 -10^3$.

This voxel anisotropy causes the number of available Fourier modes to differ between the transverse and line-of-sight directions. This can introduce artificial features in the spherically averaged power spectrum, because different $|k|$ bins probe Fourier modes with different relative contributions from transverse and line-of-sight directions. In this work, however, for simplicity and for consistency with previous LIM studies, we compute the three-dimensional power spectrum in terms of the spherically averaged $P(|k|)$, and instead account for the direction-dependent mode sampling when estimating the statistical uncertainties (see Section~\ref{sec:power_spectrum}).

\bsp	
\label{lastpage}
\end{document}